\newif\ifmnras
\mnrastrue
\ifmnras
	\documentclass[useAMS,usenatbib,usedcolumn]{mnras}
\else
	\documentclass[apj,numberedappendix]{emulateapj}
\fi
\usepackage{times}
\usepackage{graphics,epsf}
\usepackage{amsmath}                
\usepackage{amsfonts}               
\usepackage{amssymb}                
\usepackage{epsfig}                 
\usepackage{graphicx}
\usepackage{url}
\usepackage{tabularx}
\usepackage{booktabs}
\usepackage{hyperref}

\def \cm{~\mathrm{cm}}
\def \s{~\mathrm{s}}
\def \ms{~\mathrm{ms}}

\def \km{~\mathrm{km}}
\def \kms{~\mathrm{km}~\mathrm{s}^{-1}}

\def \grcc{~\mathrm{g}~\mathrm{cm}^{-3}}

\def \erg{~\mathrm{erg}}
\def \foe{~\mathrm{foe}}

\def \MeV{~\mathrm{MeV}}

\def \zams{\mathrm{ZAMS}}
\def \tpb{t_\mathrm{pb}}
\def \Ye{Y_\mathrm{e}}
\usepackage{color}

\ifmnras
	\def \aap{A\&A}
	\def \araa{ARA\&A}
	\def \arnps{Annu.~Rev.~Nucl.~Part.~Sci.}
	\def \apj{ApJ}
	\def \apjl{ApJ}
	\def \apjs{ApJS}
	\def \nat{Nature}
	\def \mnras{MNRAS}
	
	\def \prd{Phys.~Rev.~D}
	\def \prl{Phys.~Rev.~Lett.}
\fi

\def \newa{NewA}
\def \natph{Nature Physics}
\def \sci{Science}

\ifmnras
	\title[Core collapse of rotating massive star]{Asymmetric core collapse of rapidly rotating massive star}
	\author[A. Gilkis]{Avishai Gilkis \\
	Institute of Astronomy, University of Cambridge, Madingley Rise, Cambridge, CB3 0HA, UK\\
	agilkis@ast.cam.ac.uk}
\fi

\begin{document}

\ifmnras
	\pagerange{\pageref{firstpage}--\pageref{lastpage}} \pubyear{2017}

	\maketitle
\else
	\title{Asymmetric core collapse of rapidly rotating massive star}

	\author{Avishai Gilkis}
	\affil{Institute of Astronomy, University of Cambridge, Madingley Rise, Cambridge, CB3 0HA, UK\\
	agilkis@ast.cam.ac.uk}
\fi

\label{firstpage}

\begin{abstract}
Non-axisymmetric features are found in the core collapse of a rapidly rotating massive star,
which might have important implications for magnetic field amplification and production of a bipolar outflow that can explode the star,
as well as for \textit{r}-process nucleosynthesis and natal kicks.
The collapse of an evolved rapidly rotating $M_\mathrm{ZAMS}=54M_{\odot}$
star is followed in three-dimensional hydrodynamic simulations using the \textsc{flash} code with neutrino leakage.
A rotating proto-neutron star (PNS) forms with a non-zero linear velocity.
This can contribute to the natal kick of the remnant compact object.
The PNS is surrounded by a turbulent medium,
where high shearing is likely to amplify magnetic fields, which in turn can drive a bipolar outflow.
Neutron-rich material in the PNS vicinity might induce strong \textit{r}-process nucleosynthesis.
The rapidly rotating PNS possesses a rotational energy of $E_\mathrm{rot} \ga 10^{52} \erg$.
Magnetar formation proceeding in a similar fashion will be able to deposit a portion of this energy
later on in the SN ejecta through a spin down mechanism.
These processes can be important for rare supernovae generated by rapidly rotating progenitors,
even though a complete explosion is not simulated in the present study.

\end{abstract}

\section{INTRODUCTION}
\label{sec:introduction}

Most core-collapse supernovae (CCSNe) are observed to have explosion energies around
$E_\mathrm{expl} \approx 10^{51} \erg \equiv 1 \foe$ 
(e.g., \citealt{Kasen2009,Drout2011}).
In recent years, several superluminous supernovae (SLSNe) have been observed (e.g., \citealt{Quimby2011,Quimby2013,GalYam2012,Prajs2017}).
Some of these are Type IIn SNe,
and their extreme luminosity is generally attributed to the collision of the ejecta with circumstellar material (e.g., \citealt{Ofek2007,Rest2011})
ejected in a pre-explosion outburst (PEO).
The physical mechanism and the relative rarity of PEOs are yet to be fully understood
(for recent ideas see \citealt{Quataert2012,Shiode2014,Mcley2014,SokerGilkis2017}).

The case of hydrogen-poor SLSNe (Type I SLSNe, or SLSNe-I) is more complicated,
with extreme examples such as
{SN 2010ay} \citep{Sanders2012}
and {ASASSN-15lh} \citep{Dong2016}.
A variety of proposed mechanisms includes interaction with pre-explosion ejecta from pulsational pair-instability \citep{Chatzopoulos2012},
injection of energy from a millisecond magnetar (e.g., \citealt{Kasen2010,Inserra2013,Kasen2016,Sukhbold2016}),
the transition of a neutron star (NS) into a quark star (quark novae; \citealt{Ouyed2015,Ouyed2016})
and energy deposition by bipolar jets \citep{Gilkis2016}.
A combination of circumstellar interaction and magnetar spin-down has been proposed as well \citep{Chatzopoulos2016},
and magnetar birth can also be accompanied by jets \citep{Soker2016}.

Long-duration gamma-ray bursts (LGRBs) are clearly associated with broad-lined Type Ic SNe
\citep{Galama1998,WoosleyBloom2006,Modjaz2008},
and indirectly connected to SLSNe-I through the properties of their host galaxies,
cluing into the operating mechanism of these transients.
The hosts of these types of transients are low-luminosity and low-metallicity galaxies with high star formation rates \citep{Lunnan2014},
although some differences are observed \citep{Leloudas2015,Angus2016}.
It is suggested that there is a common process leading to these separate phenomena.
\cite{Metzger2015}, for example, suggest that magnetars are the powering mechanism in both SLSNe-I and LGRBs (see also \citealt{Yu2017}).
Another possibility is that jets launched from an accretion disc around a compact object power these events \citep{Woosley1993,Milosavljevic2012,Dexter2013}.

The mechanisms mentioned above call for a rapid rotation of the progenitor star,
and it is important to understand how this can be accommodated by the evolution of massive stars (see review by \citealt{Langer2012}).
The rarity of SLSNe-I and LGRBs is compatible with the requirement for high rotation,
as stellar evolution models predict that massive stars will lose most of their angular momentum \citep{Meynet2000}.
Specific binary interactions
(e.g., \citealt{Izzardetal2004,Podsiadlowskietal2004,FryerHeger2005,Cantiello2007,Yoonetal2010,deMink2013})
or chemically homogeneous stellar evolution \citep{Yoon2005,Woosley2006,Martins2013,MandelanddeMink2016,Song2016}
are required to supply the high rotation needed for extreme explosion scenarios.

Recent three-dimensional hydrodynamic core-collapse simulations of rotating stars have yielded intriguing results.
\cite{Nakamura2014} and \cite{Takiwaki2016} showed a preferred explosion direction \textit{perpendicular} to the rotation axis.
\cite{Mosta2014} preformed three-dimensional magnetorotational simulations of core collapse,
finding a phenomenon of magnetically-inflated asymmetric lobes.
Other studies focused on the properties of gravitational waves expected from the collapse of a rotating star \citep{Ott2007,Kuroda2014}.
\cite{Iwakami2009} imposed rotation after core collapse,
and their findings suggest that rotation can affect the standing accretion-shock instability
(SASI; e.g., \citealt{BlondinMezzacappa2003, BlondinMezzacappa2007, Fernandez2010}).

To advance the understanding of the possible origin and mechanism of SLSNe-I,
atypical progenitor stars should be considered,
such as those with extremely rapid rotation rates, or very high mass.
In the present study I explore the properties of the post-collapse flow in a very massive rapidly rotating star.
Three-dimensional hydrodynamic simulation of the stellar collapse are performed,
including deleptonization and a neutrino leakage scheme for heating and cooling.
For comparison, a slowly rotating case is simulated as well.
In section \ref{sec:setup} I describe the numerical setup and method.
The flow structure of the collapse and implications for CCSNe are presented in section \ref{sec:results}.
I summarize in section \ref{sec:summary}.


\section{NUMERICAL SETUP}
\label{sec:setup}

\subsection{Progenitor modeling}
\label{subsec:mesa}

A stellar model constructed by 
Modules for Experiments in Stellar Astrophysics (\textsc{mesa} version 7624; \citealt{Paxton2011,Paxton2013,Paxton2015})
is used,
with an initial mass of $M_\zams=54 M_\odot$ and metallicity of $Z=0.014$.
Due to stellar winds calculated here with the so-called `Dutch' scheme (e.g., \citealt{Nugis2000,Vink2001})
the final mass is $20.5M_\odot$.
The choice of an atypically high initial mass is motivated by the pursuit of models for rare highly-energetic events.
The study of rapidly rotating CCSN progenitor stars of lower masses  is deferred to a future paper.

The initial rotation is $0.55$ of the breakup value,
which corresponds to a surface rotation velocity of $v_\zams=360\kms$.
Magnetic braking by the Spruit-Tayler dynamo \citep{Spruit2002} is neglected,
effectively resulting in a high core rotation which might not be expected in a single-star model.
The pre-collapse core rotation is rather insensitive to the initial rotation rate,
and the decisive factor is the Spruit-Tayler dynamo (see Appendix \ref{app:corej}).
At this stage, the star is a Wolf-Rayet star,
with an effective surface temperature of $T=2.4 \times 10^5 K$,
photospheric radius of $R=0.55R_\odot$
and luminosity of $L=9 \times 10^5 L_\odot$.

\begin{figure}
    \includegraphics*[scale=0.53]{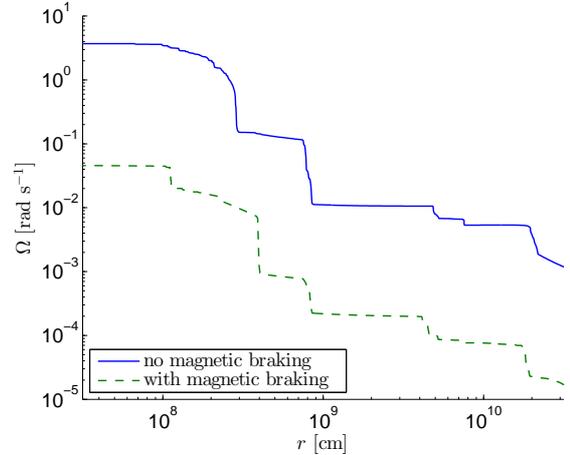} \\
\caption{Comparing angular velocity of two stellar models with same initial conditions, just before core collapse,
where in one model magnetic braking is included and in the other it is neglected. 
$r$ is the shell radius (distance from the centre of the star),
and $\Omega$ is assumed constant within each shell.}
      \label{fig:omega}
\end{figure}

\begin{figure}
\begin{tabular}{c}
    \includegraphics*[scale=0.53]{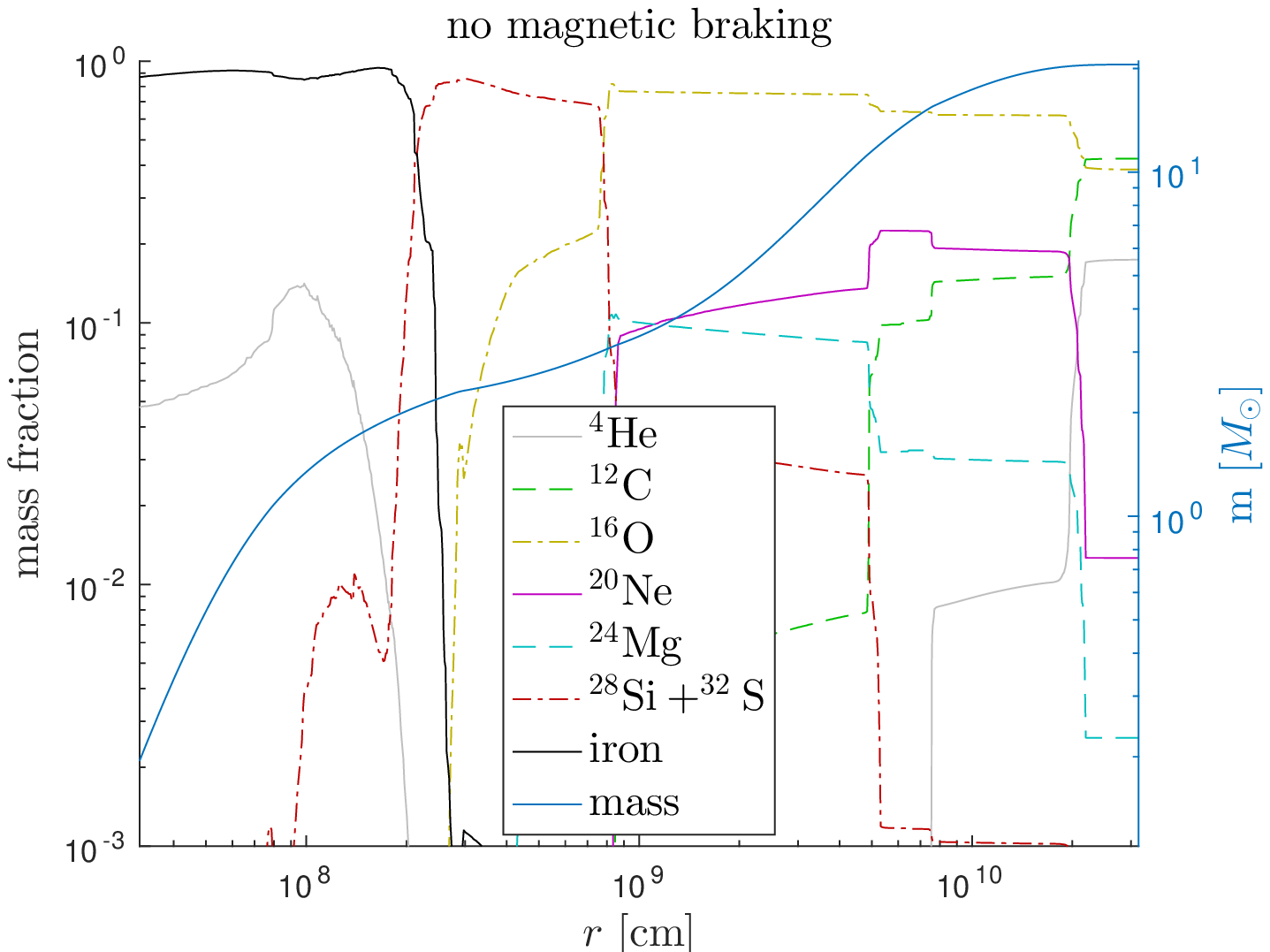} \\
    \includegraphics*[scale=0.53]{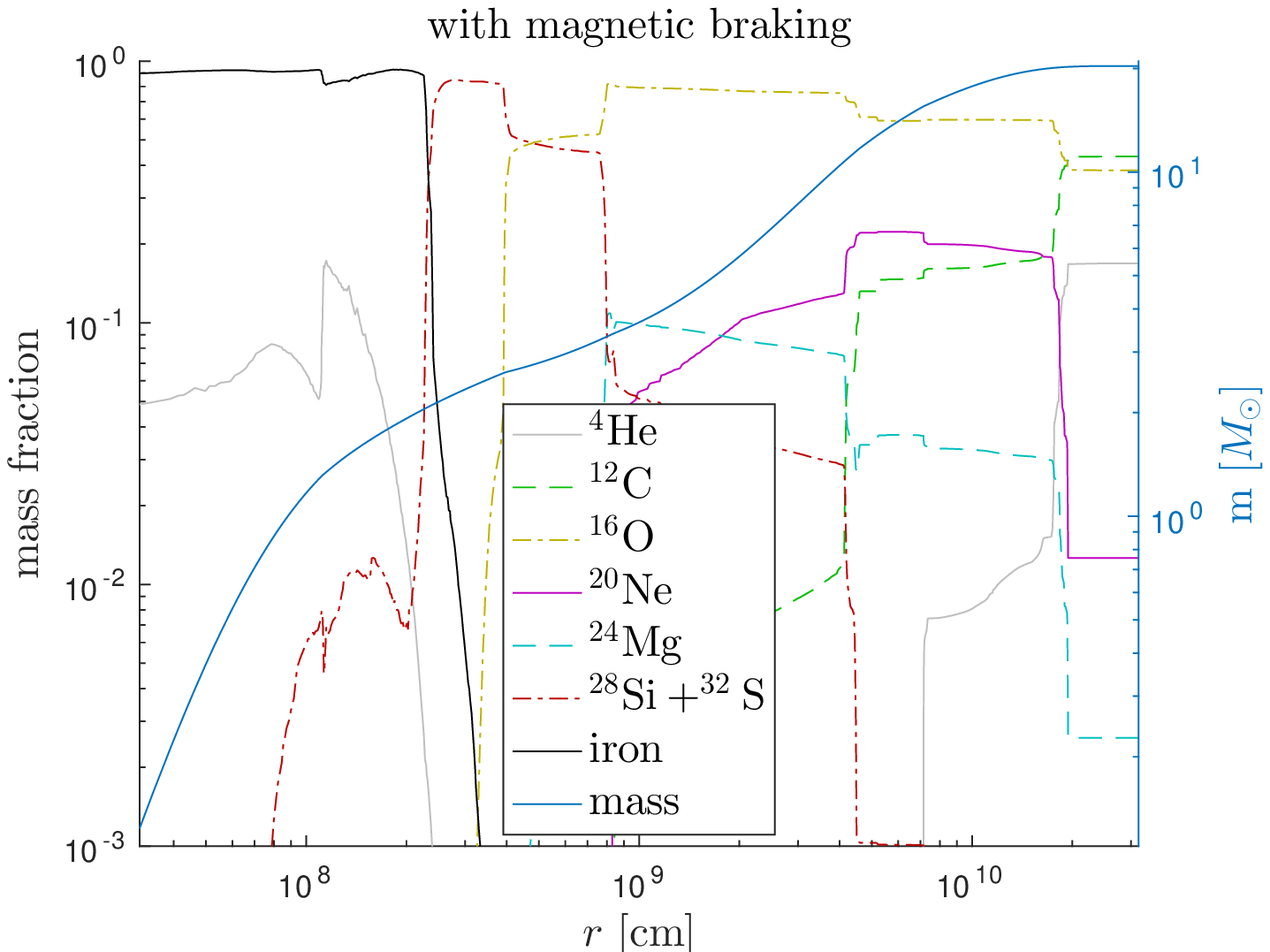} \\
\end{tabular}
      \caption{Detailed composition for the stellar models just before core collapse
      (when the maximal inward velocity reaches $v = 1000 \kms$)
      of a rotating $M_\zams=54 M_\odot$ star with metallicity of $Z=0.014$,
      with magnetic braking neglected (top), or included (bottom).
      The stellar parameters at this stage are detailed in Table \ref{tab:models}.}
      \label{fig:composition}
\end{figure}

A second stellar model is evolved with magnetic braking included.
The inclusion of magnetic braking
results in a core rotation lower by almost two orders of magnitude
(Fig. \ref{fig:omega}).
The stellar model which includes magnetic braking
has a pre-collapse total rotational kinetic energy of $E_\mathrm{rot,slow} = 2.07 \times 10^{46} \erg$.
For the case where magnetic braking was not included,
the total rotational kinetic energy is $E_\mathrm{rot,fast} = 1.58 \times 10^{50} \erg$.
The iron core of the fast rotator has a mass of $M_\mathrm{iron,fast} \approx 2 M_\odot$
(almost the same as with magnetic braking)
and rotational kinetic energy of $E_\mathrm{rot,iron,fast} = 1.07 \times 10^{50} \erg$ --
a significant fraction of the entire rotational energy of the stellar model.
The high initial rotation rate considerably affects the stellar evolution,
even while on the main sequence (MS),
compared with a modelled star of the same initial mass and composition but no rotation at all (this model is not presented here).
The differences between the two rotating models, differing by the process of magnetic braking,
are much less pronounced and start to appear only at late post-MS evolutionary stages.
The detailed composition of the stellar models is shown in Fig. \ref{fig:composition},
and the main parameters of the two models are summarized in Table \ref{tab:models}.
\begin{table}
 \caption{Properties of the stellar models just before core collapse.}
 \label{tab:models}
 \begin{tabular}{lll}
  \hline
  Spruit dynamo & included & neglected\\[3pt]
  \hline
  Luminosity [$L_\odot$] & $9.3 \times 10^5$ & $9.1 \times 10^5$\\[3pt]
  Effective surface temperature [$\mathrm{K}$] & $2.5 \times 10^5$ & $2.4 \times 10^5$\\[3pt]
  Total mass [$M_\odot$] & $20.3$ & $20.5$\\[3pt]
  Photosphere radius [$R_\odot$] & $0.5$ & $0.55$\\[3pt]
  Iron core mass [$M_\odot$] & $2.01$ & $1.96$\\[3pt]
  Iron core radius [$\mathrm{km}$] & $2187$ & $1858$\\[3pt]
  \hline
 \end{tabular}
\end{table}

The stellar models are followed until a core of iron-group elements forms,
and then the core becomes unstable and starts to collapse.
The \textsc{mesa} evolution is terminated when the maximal infall velocity reaches $v = 1000 \kms$.
At this stage the one-dimensional profile is mapped into the three-dimensional grid of a hydrodynamic simulation,
described in the next section.

\subsection{Hydrodynamic simulations}
\label{subsec:flash}

The non-relativistic hydrodynamic equations are solved with version 4.3 of the widely used code \textsc{flash} \citep{Fryxell2000},
employing a Cartesian grid for three-dimensional simulations.
Newtonian self-gravity is included employing the spherical multipole approximation of \cite{Couchetal2013} for solving Poisson's equation,
with a multipole cutoff of $l_\mathrm{max}=16$.
The equation of state of \cite{LattimerSwesty1991} is employed with incompressibility parameter of $K=220\MeV$.

The density ($\rho$), temperature ($T$), electron fraction ($\Ye$) and radial velocity ($v_r$)
of the one-dimensional \textsc{mesa} profile are mapped into the three-dimensional grid.
The small deformations of the one-dimensional shells due to rotation are neglected,
so that the initial profiles of $\rho$, $T$, and $\Ye$ are spherically symmetric.
A non-radial velocity component is added according to the angular velocity (Fig. \ref{fig:omega}),
with a constant angular velocity for each radial location, $\Omega\left(r\right)$.
The added velocity component in the plane perpendicular to the rotation axis (the $xz$ plane)
is then
$v_{xz}=\sqrt{x^2+z^2}\Omega\left(r\right)$.

The three-dimensional simulation domain is a cube with edge length of $12000 \km$
centred around the star.
The mass enclosed in this domain is about $3 M_\odot$,
out of the total stellar mass of $20.5M_\odot$.
For studying the early time post-bounce flow dynamics this is sufficient.
With nine levels of adaptive mesh refinement the finest resolution of the grid is $1.95 \km$.

\subsection{Neutrino physics}
\label{subsec:neutrino}

Deleptonization of the core, approximate treatment of neutrino transport by a leakage scheme and heating by neutrinos
are treated according to the methods of \cite{OConnor2010}.
The usage in the \textsc{flash} code is the implementation described in \cite{CouchOConnor2014},
and here only the main aspects are repeated.

Up until bounce,
defined as the first time when the maximal density exceeds $2\times 10^{14} \grcc$
and the maximal entropy per nucleon is above $3 k_\mathrm{B}$,
core deleptonization is according to a fit of the electron fraction as function of density from the 1D simulations of
\cite{Liebendorfer2005}.
As hypothesized by \cite{Liebendorfer2005},
the application of the scheme to cases of fast rotation (as in the present study)
should be a reasonable approximation,
although this should be checked in the future.
After core bounce,
further deleptonization is according to the leakage scheme.

The leakage scheme interpolates between 
the optically thick and the optically thin regimes,
where in the former neutrinos `leak' out on a diffusion timescale,
and immediately in the latter.
Effective lepton and energy emission rates are obtained by interpolating between the diffusion rate and `free' emission rate.
The diffusion rates require a calculation of the local optical depth.
This is done on a spherical grid of radial rays,
where the computation of the optical depth by radial integration from infinity is simple,
and then interpolated back to the Cartesian grid of the hydrodynamic simulation.
The spherical grid has 37 polar divisions
(from $\theta=0$ to $\theta=\pi$, where $\theta$ is the angle relative to the rotation axis $y$)
and 75 azimuthal divisions
(from $\phi=0$ to $\phi=2\pi$, where $\phi$ is the angle from the $x$ axis in the $xz$ plane).
Each ray consists of 1000 radial zones.
The radial rays are uniformly spaced from $r=0.5 \km$
up to $r=150 \km$,
followed by a logarithmic distribution up to $r=3000 \km$ (where the stellar matter is virtually transparent to neutrinos).

Local neutrino heating
is according to equation (2) of \cite{CouchOConnor2014},
which depends on the local nucleon number density,
the mean squared energy of the neutrinos at the neutrinosphere,
and the neutrino flux integrated along a ray from the centre (including flux reduction by heating along the way).
This heating is then subtracted from the neutrino flux, affecting regions farther out along the radial ray.
The interactions taken into account in the portrayed scheme are the following.
For the free emission the included interactions are
electron and positron capture on protons and neutrons, respectively,
as well as thermal processes.
The opacity calculations include absorption of electron neutrinos on neutrons and electron antineutrinos on protons,
and elastic scattering of all flavors of neutrinos on protons and neutrons.
Finally, charged-current heating by absorption of electron neutrinos on neutrons and antineutrinos on protons is included.
These reactions are the most relevant, although processes of lesser importance such as neutrino-electron scattering can also have an influence (e.g., \citealt{Lentz2012}), such as reduction of the average neutrino energy and accordingly the opacity to neutrinos.
Further details of the emplyed neutrino physics can be found in \cite{OConnor2010}, \cite{Ott2013} and \cite{CouchOConnor2014}, and references therein.

\section{RESULTS}
\label{sec:results}

\subsection{Flow structure}
\label{subsec:structure}

The two simulations were run in \textsc{flash} for $600 \ms$,
corresponding to a final post-bounce time of $\tpb = 384 \ms$ for the fast rotator model,
and $\tpb = 323 \ms$ for the slow rotator.
The main features of interest found for the fast rotator model are not seen in the slow rotator,
and the slow rotator behaves in a qualitatively similar fashion to a non-rotating progenitor,
forming an approximately spherical stalled accretion shock for a time of $\ga 150 \ms$.
As there are extensive detailed studies of non-rotating core collapse with a more accurate description of neutrino physics and higher resolution
(e.g., \citealt{Lentz2015,Melson2015,Kuroda2016,Roberts2016} for recent works;
\citealt{Janka2012}, \citealt{Janka2016} and \citealt{Muller2016} for reviews),
this will not be the focus of this paper.

\begin{figure*}
\begin{tabular}{cc}
    \includegraphics*[scale=0.22]{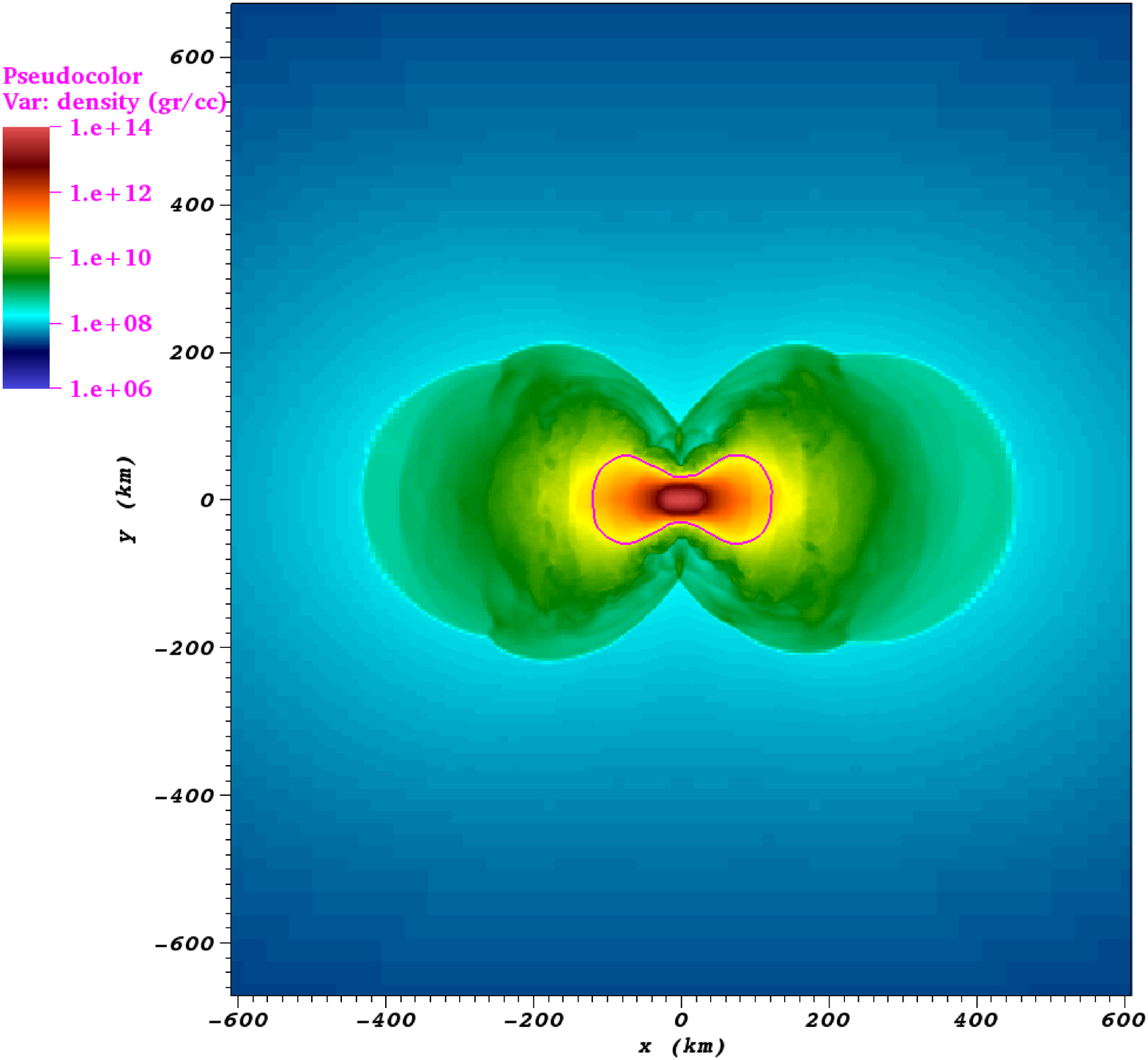} &
    \includegraphics*[scale=0.22]{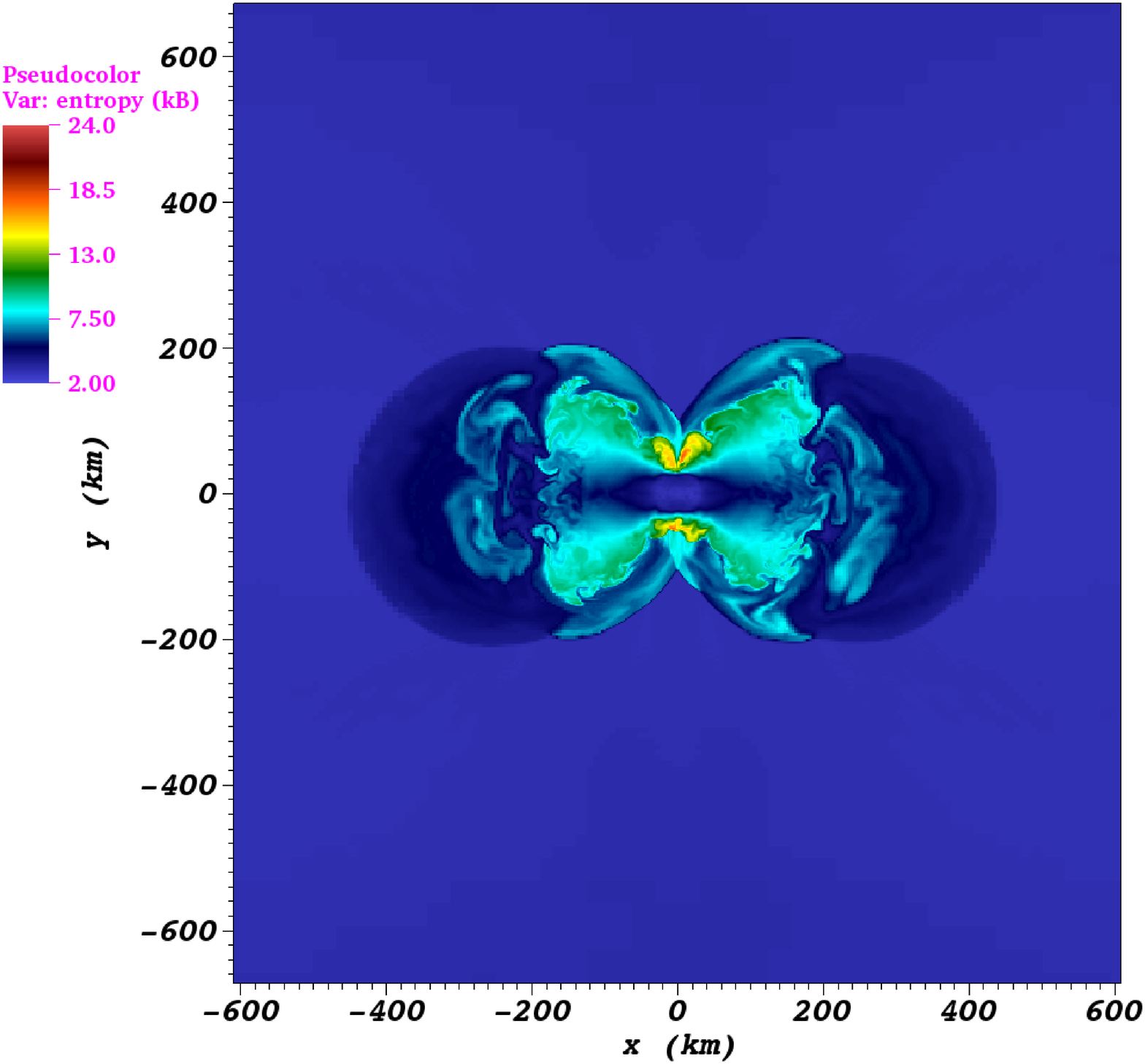} \\
    \includegraphics*[scale=0.22]{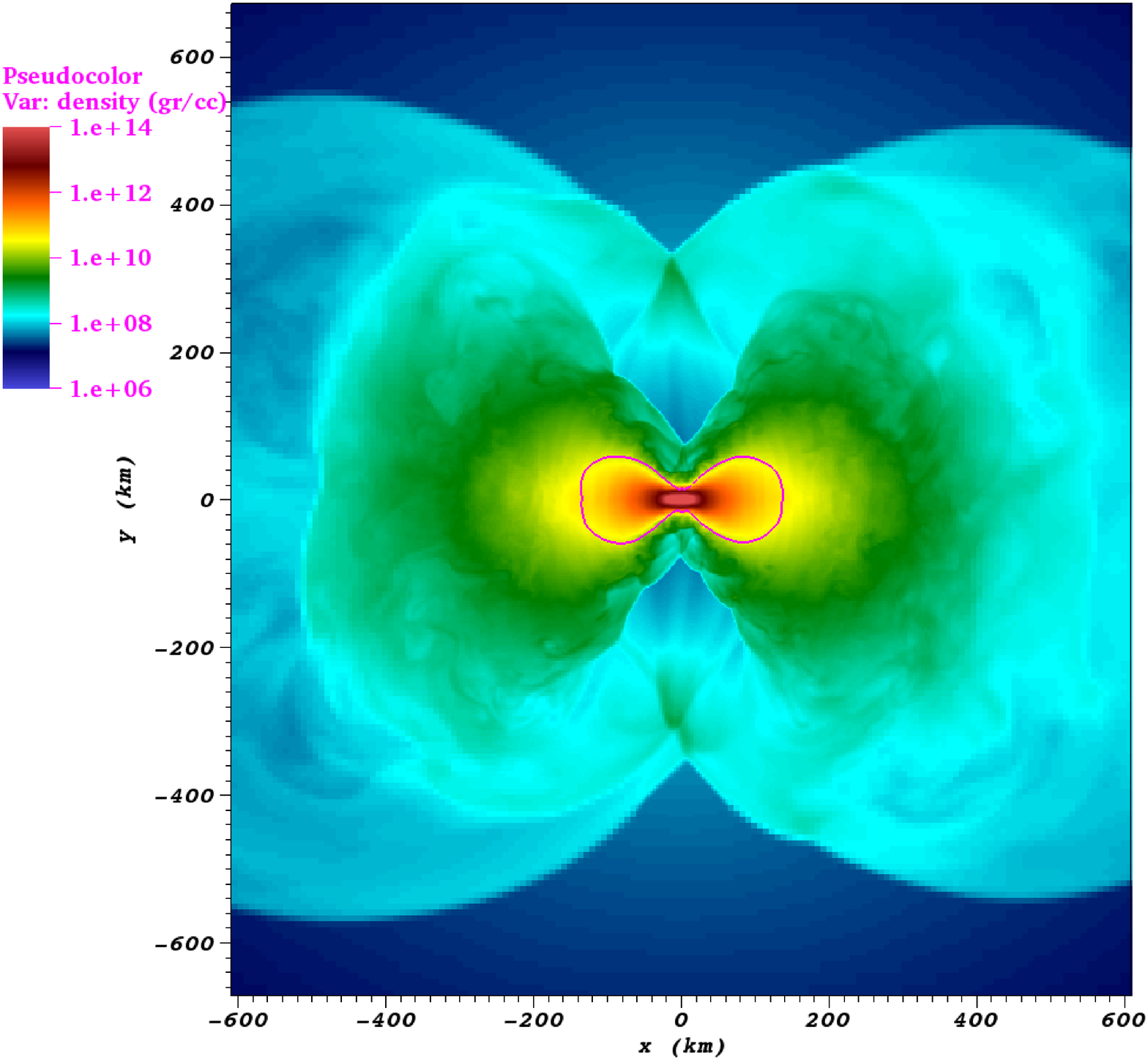} &
    \includegraphics*[scale=0.22]{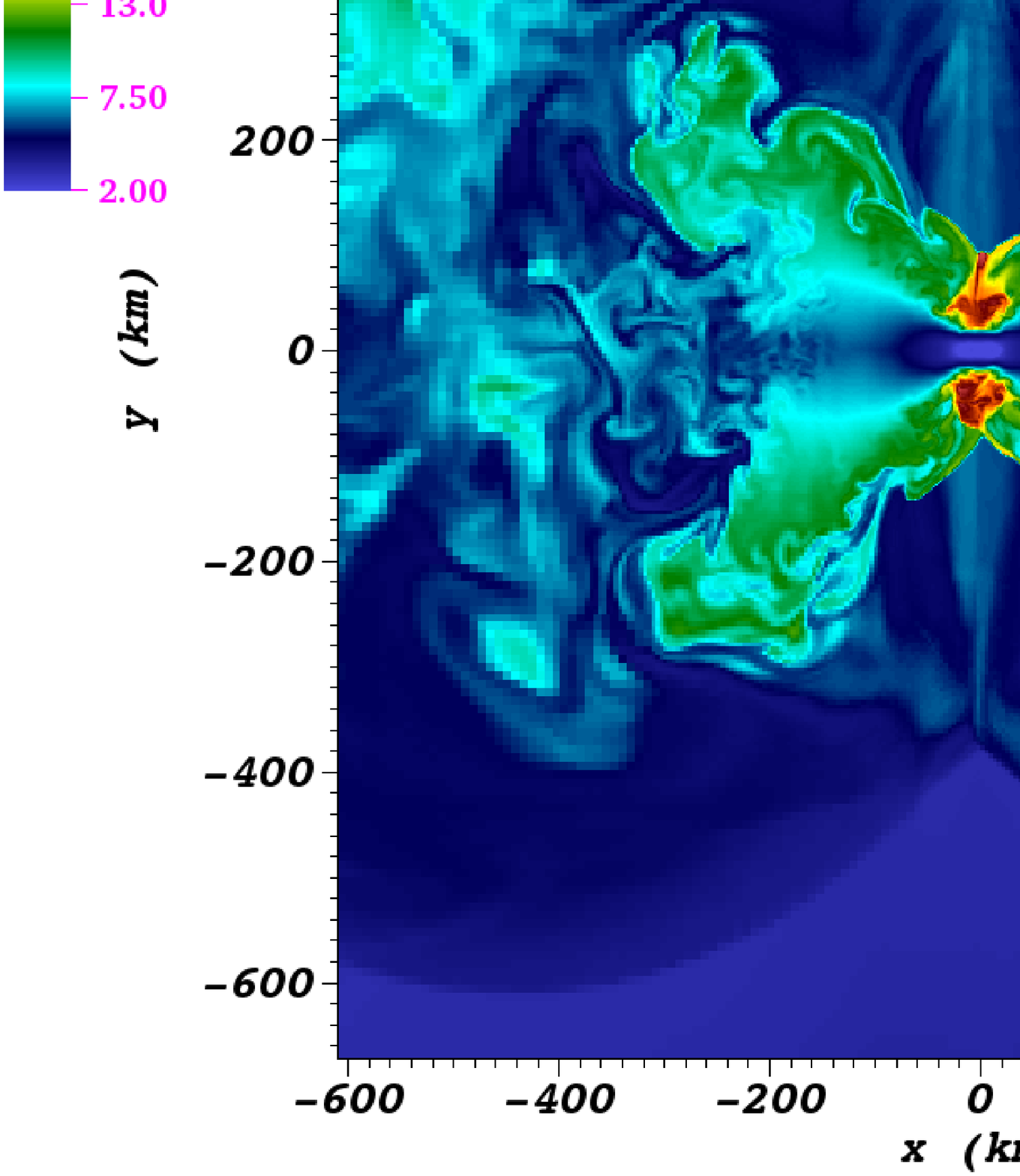} \\
\end{tabular}
      \caption{Density (left) and entropy (right) colour maps in the $xy$ plane,
      where $y$ is the axis of rotation.
      The density colour coding runs from 
      $\rho=10^6 \grcc$ to $\rho=10^{14} \grcc$.
      Thick purple contours superimposed on the density colour maps indicate the neutrinosphere at the presented simulation times,
      where the optical depth of electron neutrinos is $\tau_{\nu_\mathrm{e}}=2/3$.
      Entropy is shown in units of $k_\mathrm{B}$ per nucleon.
      The presented times are $\tpb=104\ms$ (top panels)
      and $\tpb=224\ms$ (bottom panels).}
      \label{fig:flow}
\end{figure*}

Fig. \ref{fig:flow} shows the properties of the flow structure at two times after core bounce
for the rapidly rotating model, $\tpb=104\ms$ and $\tpb=224\ms$.
A torus-like dense region is clearly seen surrounding the oblate proto-neutron star (PNS).
Above and below the PNS are high-entropy turbulent regions.
When the two `polar holes' (one at each side of the equatorial plane) in the torus are closed,
the shock waves produce high-entropy regions near the rotation axis.

As the PNS is oblate, the neutrino emission is anisotropic, and the neutrinosphere is accordingly not spherical.
This introduces a difficulty with the employment of radial rays for the neutrino transport,
as the neutrinosphere morphology obviously implies that there should be significant non-radial neutrino propagation.
Future studies will have to address this with better treatment of the neutrino physics.

\begin{figure}
\begin{tabular}{c}
    \includegraphics*[scale=0.22]{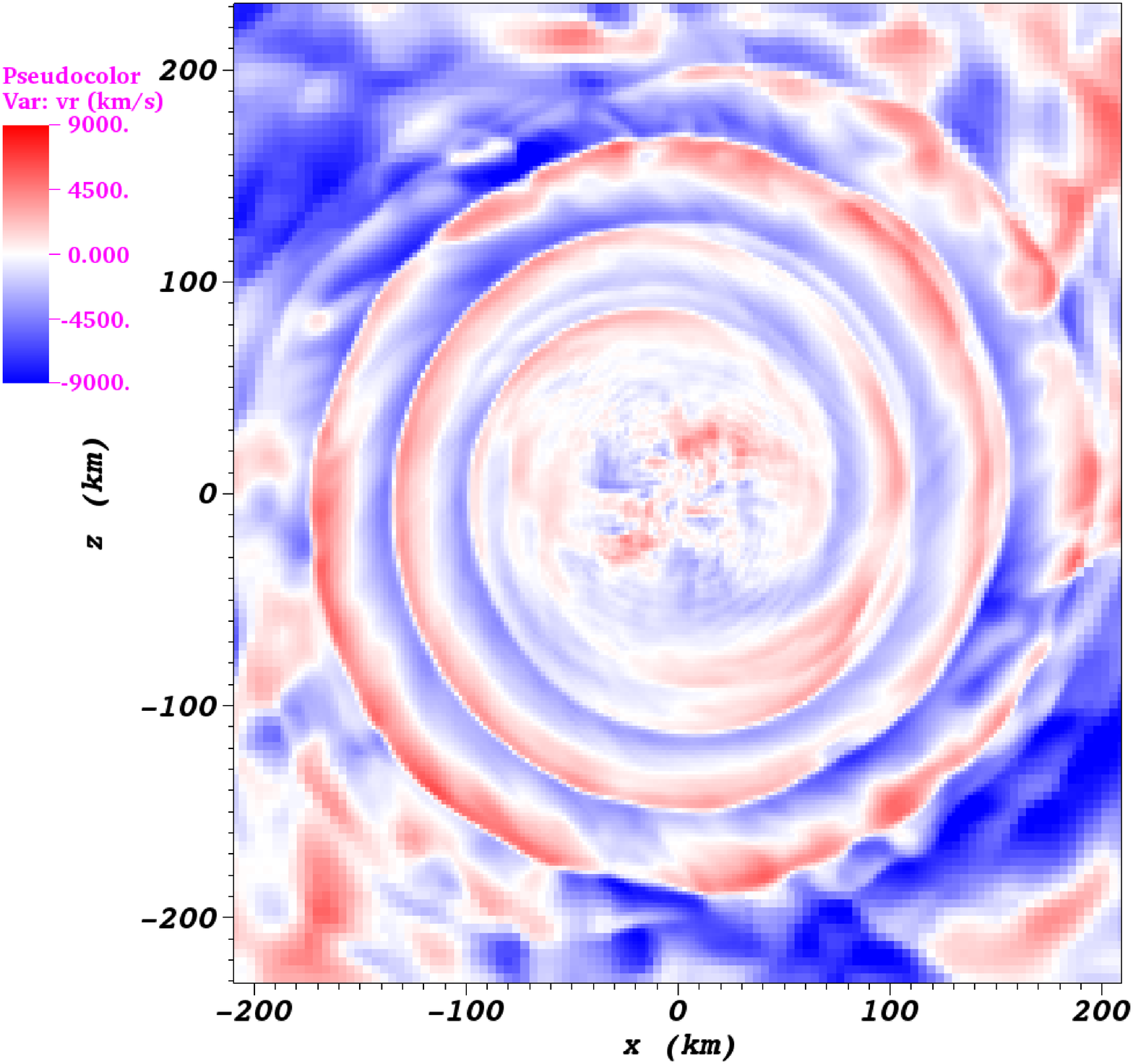} \\
    \includegraphics*[scale=0.22]{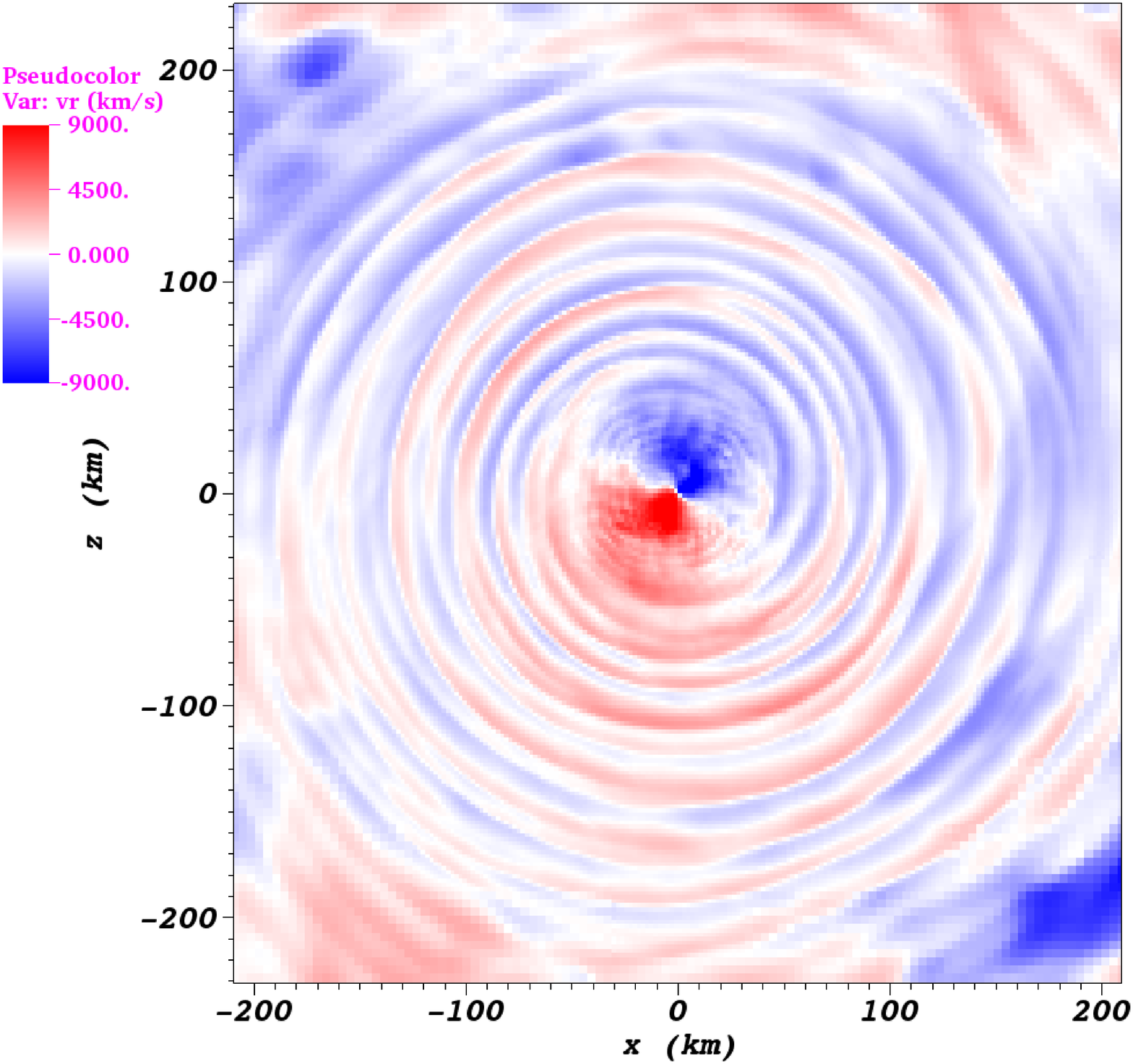} \\
\end{tabular}
      \caption{Radial velocity in the equatorial ($xz$) plane of the star at two times after core bounce,
      $\tpb = 104 \ms$  (top),
      and $\tpb = 224 \ms$  (bottom).
      The colour coding runs from in-fall of $v_r=-9000\kms$ (blue)
      to outflow of $v_r=9000\kms$ (red).
      The effective PNS radius calculated from its volume is $r_\mathrm{eff,PNS}\approx 75 \km$,
      but as the PNS is highly oblate,
      the radius of its cross section in the $xz$ plane is $r_\mathrm{eq,PNS}\approx 100 \km$.
      The equatorial Keplerian velocity at its surface is $r_\mathrm{Kep,PNS}\approx 47500 \kms$.}
      \label{fig:radialvelocityspiral}
\end{figure}

Fig. \ref{fig:radialvelocityspiral} shows the radial velocity in the equatorial plane of the star.
At the earlier time shown, a spiral pattern is seen,
and at the later time ripples of inward and outward radial velocity surround the central region.
These phenomena are strongly influenced by the self-gravity of the gas.
The ratio between the gravitational time, defined as $t_\mathrm{G}=\left(G\rho\right)^{-1/2}$,
to the dynamical time $t_\mathrm{dyn}=r/c_s$,
is very close to unity up to around $r\approx 150-200\km$ near the equatorial plane.
This is the region in which the spiral pattern and ripples are seen.
Spiral patterns forming in the collapse of a rotating massive star have been reported also by \cite{Takiwaki2016},
and have been suggested to drive circular polarizations of gravitational waves emitted from CCSNe \citep{Hayama2016}.

\begin{figure}
    \includegraphics*[scale=0.53]{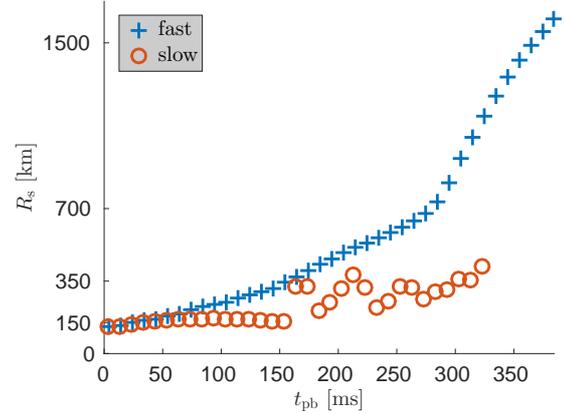} \\
\caption{Shock radius calculated from the volume contained within the surface on which the entropy per nucleon is $6k_\mathrm{B}$,
as function of time after core bounce for the rapidly rotating progenitor (pluses) and for the slow rotator (circles).}
      \label{fig:shock}
\end{figure}

At the end of the simulation, the outer regions of the star still experiences ongoing collapse,
while an outgoing shock wave is propagating in the inner part.
Fig. \ref{fig:shock} shows the calculated shock radius evolution.
While an outgoing shock is seen for both models at the end of the simulation,
the diagnostic energy\footnote{`Diagnostic energy' is defined here as the sum of kinetic, internal and gravitational energy for regions where the energy sum and the radial velocity are both positive.}
is $E_\mathrm{diag}\simeq 0.9 \foe$ and $E_\mathrm{diag}\simeq 0.02 \foe$ for the rapidly- and slowly rotating progenitors, respectively.
The binding energy\footnote{The gravitational binding energy is computed as the absolute value of integrating $E_\mathrm{grav}+E_\mathrm{int}$ from $E_\mathrm{bind}=0$ at the surface, inwards.} of the outer shells is $E_\mathrm{bind}\approx 3 \foe$, somewhat higher than the diagnostic energy of the rapidly rotating model (although a longer simulation time is needed for a conclusive comparison), and significantly larger than the slowly rotating case (note that this is resolution sensitive; see Appendix \ref{app:res}).
This can be interpreted as a failure to explode the star, though not definitively.
Still,
key physical ingredients not taken into account in the simulation --
magnetic fields and their amplification --
might help bring about a successful CCSN.
I elaborate on the possible explosion and its attributes in the next sections.

\subsection{Implications for magnetic activity}
\label{subsec:magnetic}

\begin{figure}
\begin{tabular}{c}
    \includegraphics*[scale=0.22]{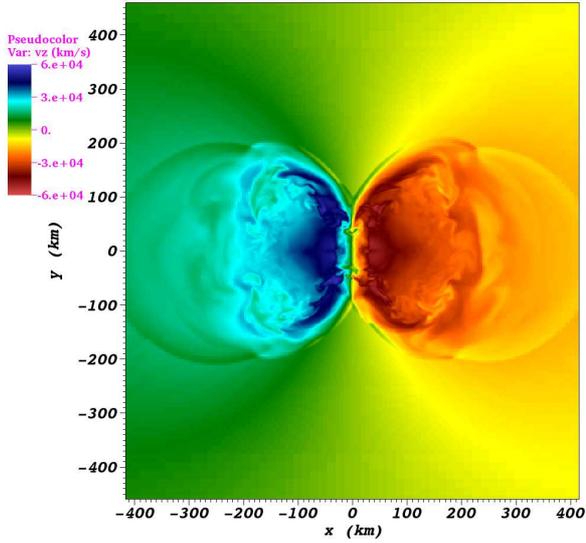} \\
    \includegraphics*[scale=0.22]{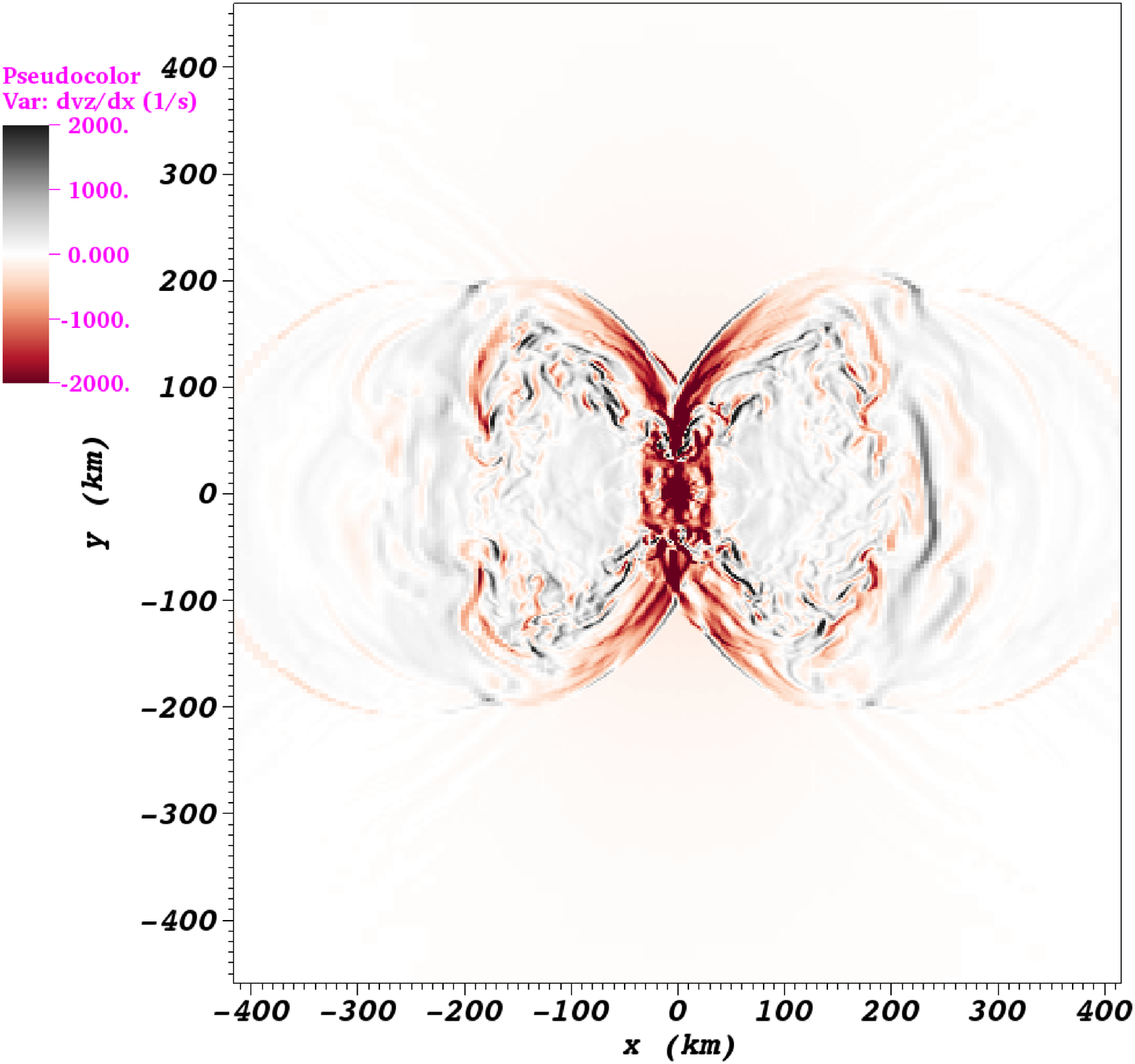} \\
\end{tabular}
	\caption{\textit{Top:} Velocity perpendicular to the $xy$ plane at $\tpb=104\ms$.
	There is a general counterclockwise rotation --
	blue hues represent material flow outwards from the paper,
	and red hues represent an inward motion
	(the $z$ axis points outwards from the paper).
	\textit{Bottom:} Derivative in the $x$ direction of the velocity shown in the top panel,
	i.e., $\frac{\partial v_z}{\partial x}$ in units of $s^{-1}$ and a scale given by the colour bar.}
      \label{fig:shear1}
\end{figure}

The velocity perpendicular to the meridional plane and its derivative within this plane with respect to the direction perpendicular to the initial rotation axis
are presented in Fig. \ref{fig:shear1} for a time of $\tpb=104\ms$.
In Fig. \ref{fig:shear1} the meridional plane is taken to be the $xy$ plane,
such that the velocity presented is in the $z$ direction , $v_z$,
and the derivative shown is $\frac{\partial v_z}{\partial x}$.
This illustrates the strong shearing present in the turbulent flow.
The regions of highest shearing are in the high entropy regions above and below
the equatorial plane (Fig. \ref{fig:flow}).
The shearing reaches quantitative values of
$\frac{\partial v_z}{\partial x} \approx 2000 \s^{-1}$.
For comparison, the Keplerian velocity divided by the distance from the rotation axis
is of an order of magnitude lower.

\cite{Mosta2015} have shown that turbulent shearing around a rapidly rotating PNS
can facilitate fast growth of magnetic fields through 
the magnetorotational instability (see also \citealt{Sawai2013}).
The amplified magnetic fields might then launch jets which can explode the star.
I suggest that the flow structure seen in the simulation of a rapidly rotating progenitor reported here is likewise favorable for a jet-driven explosion.
This will have to be confirmed by magnetohydrodynamic simulations,
which \cite{Mosta2015} show require extremely high resolution (see also \citealt{Masada2015,Rembiasz2016}).

It is important to note that the rapidly rotating model studied here is evolved without taking into account the effect of magnetic fields on the core rotation rate (see Appendix \ref{app:corej}).
This is inconsistent with the requirement of a seed magnetic field to be amplified for the generation of jet outflows.
The focus of this work, though, is the collapse dynamics of a rapidly rotating massive stellar core.
A fully self-consistent model
will have to allow for the formation of a rapidly rotating pre-collapse core in the presence of magnetic fields,
and will be explored in a future study.
Sufficiently rapid core rotation is expected for very low metallicity single stars (e.g., \citealt{Yoon2006})
or through binary interactions (e.g., \citealt{Cantiello2007}).

\subsection{PNS formation}
\label{subsec:natal}

\begin{figure}
    \includegraphics*[scale=0.53]{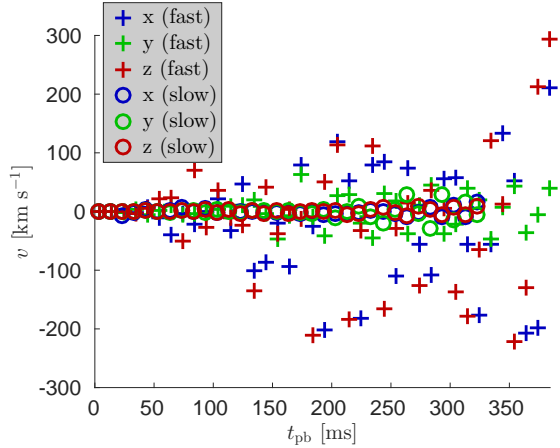} \\
      \caption{Average velocity of the PNS,
	defined as the region where the density is $\rho > 10^{11} \grcc$,
	as function of time after core bounce for the rapidly rotating progenitor (pluses)
	and for the slow rotator (circles).}
      \label{fig:kick}
\end{figure}

\begin{figure}
\begin{tabular}{c}
    \includegraphics*[scale=0.53]{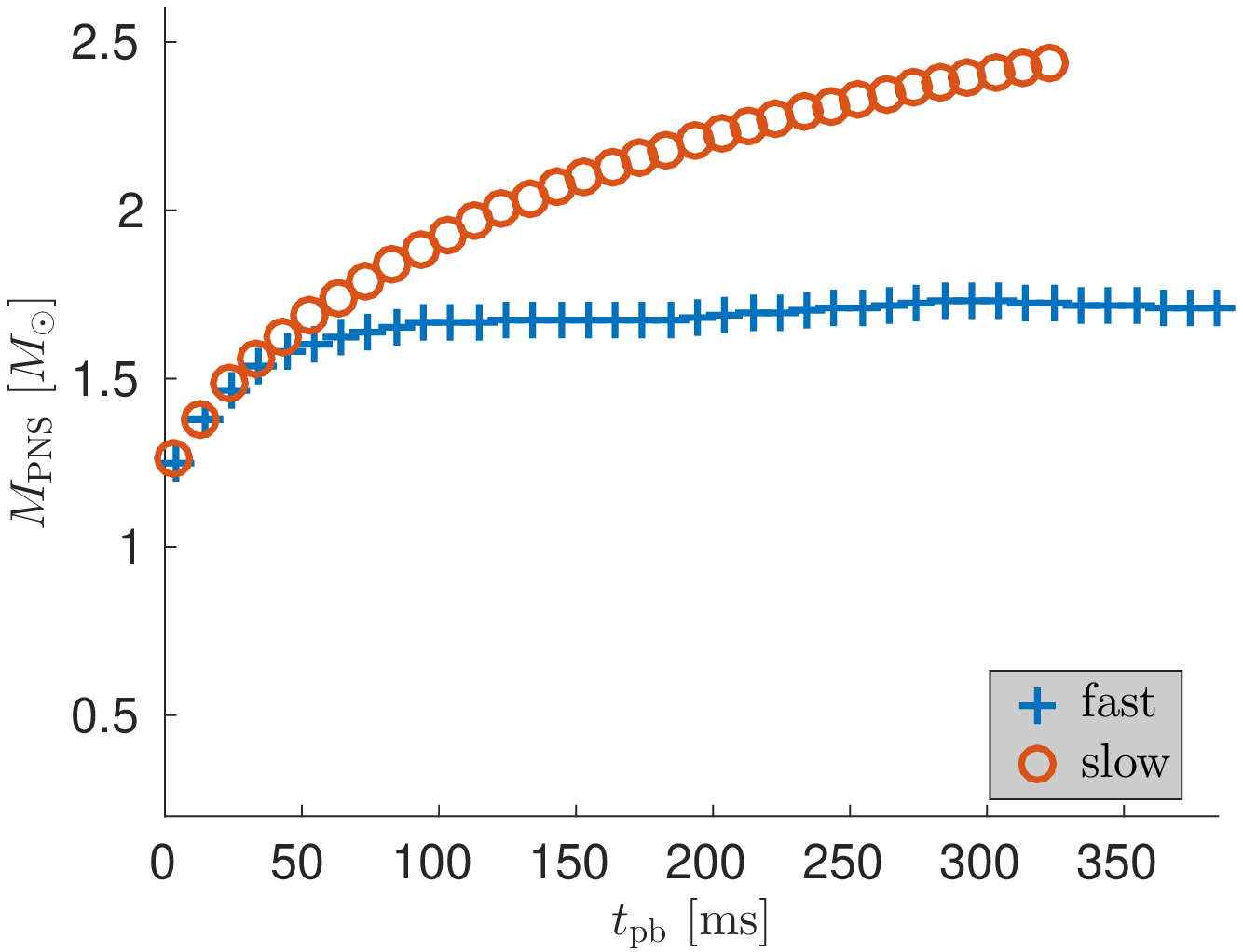} \\
    \includegraphics*[scale=0.53]{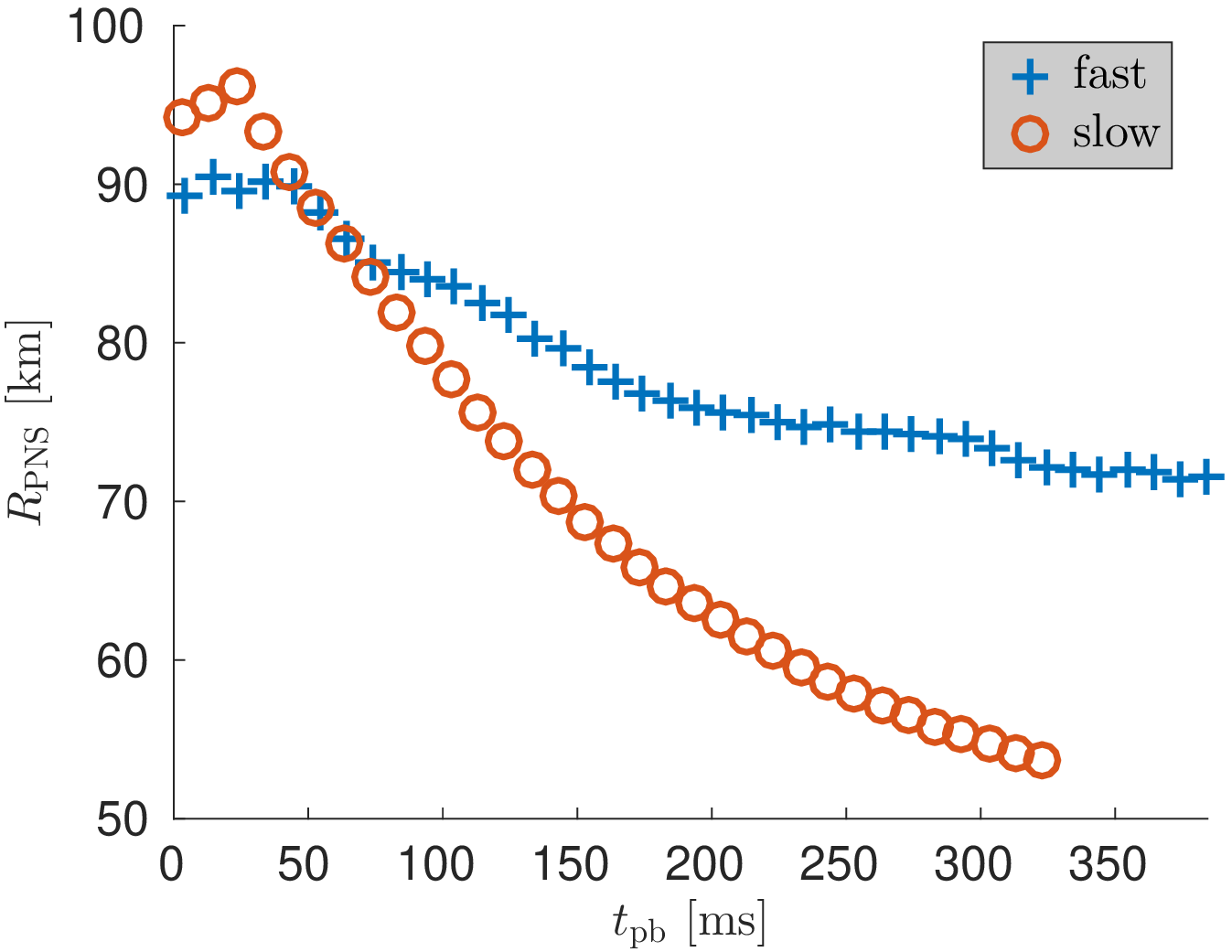} \\
    \end{tabular}
\caption{\textit{Top:} Baryonic PNS mass as function of time after core bounce for the rapidly rotating progenitor (pluses)
	and for the slow rotator (circles),
	where the PNS is defined as the region where the density is $\rho > 10^{11} \grcc$.
	\textit{Bottom:} PNS effective radius ($\pi R_\mathrm{PNS}^3=0.75V_\mathrm{PNS}$, where $V_\mathrm{PNS}$ is the PNS volume).}
      \label{fig:pns}
\end{figure}

The PNS is defined as the region in the simulation where the material density is above
$10^{11} \grcc$ (this definition is common in other works, e.g., \citealt{Nakamura2014}).
The total mass, momentum, angular momentum, and moment of inertia of the PNS are calculated after core bounce
for the simulation of a rapidly rotating progenitor, as well as for the slow rotator.
The derived linear velocity is shown in Fig. \ref{fig:kick},
and the total mass and effective radius are presented in Fig. \ref{fig:pns}.

The slow rotator acquires a very low velocity,
of $v \la 30 \kms$.
For the rapidly rotating progenitor the picture is different.
The $x$ and  $z$ components of the velocity (in the plane perpendicular to the axis of rotation)
of the fast rotator are highly variable,
fluctuating in a range of above $100 \kms$.
At the end of the simulation, the PNS velocity in the $xz$ plane is $v_{xz} \approx 360 \kms$.
The $y$ component of the velocity is somewhat less variable.
Although the fluctuations  are large,
there is no monotonous growth in the linear velocity,
and it is difficult to deduce the final kick velocity without a full explosion simulation.
Still, due to their large amplitude, these fluctuations merit discussion.

At the end of the simulations,
the mass of the PNS for the slowly rotating case exceeds $M_\mathrm{PNS} \ga 2.4 M_\odot$,
and is still growing.
It is likely that BH formation is inevitable.
For the rapid rotator the final PNS mass in the simulation is $M_\mathrm{PNS} \approx 1.7 M_\odot$.
If the accretion is not stopped by an explosion,
the PNS will continue to grow in mass, and its kick velocity will decrease.
Eventually it will collapse into a BH with low or zero kick velocity.
If, however, a bipolar outflow drives a successful explosion
as suggested in section \ref{subsec:magnetic},
the PNS will retain its high velocity.
The final kick velocity of the remnant depends on the stage at which the explosion occurs,
as well as the efficiency of expelling material,
and in this manner the kick velocity depends on the mass of the ejecta \citep{Bray2016}.
An inefficient feedback will result in continued accretion and the formation of a BH,
although a higher energy SN might be the result \citep{Gilkis2016}.
Still, at some point accretion onto the PNS (or BH) will stop,
and a BH or NS with non-zero natal kick will emerge from the explosion.
Non-zero BH kicks are favoured by some recent studies (e.g., \citealt{Repetto2012}).

As seen in Fig. \ref{fig:kick}, the kick velocity components in the equatorial plane are larger than in the direction of the rotation axis.
The turbulent nature of the gas flow from which jets are conjectured to be driven might cause asymmetric mass ejection,
further enhancing the kick velocity with an additional component in the $y$ axis by the mechanism proposed by \cite{Janka2013}.
An analysis by \cite{Repetto2015}
suggests that at least some BHs form with relatively high natal kicks
(see also \citealt{Mandel2016}).

\begin{figure}
\begin{tabular}{c}
    \includegraphics*[scale=0.53]{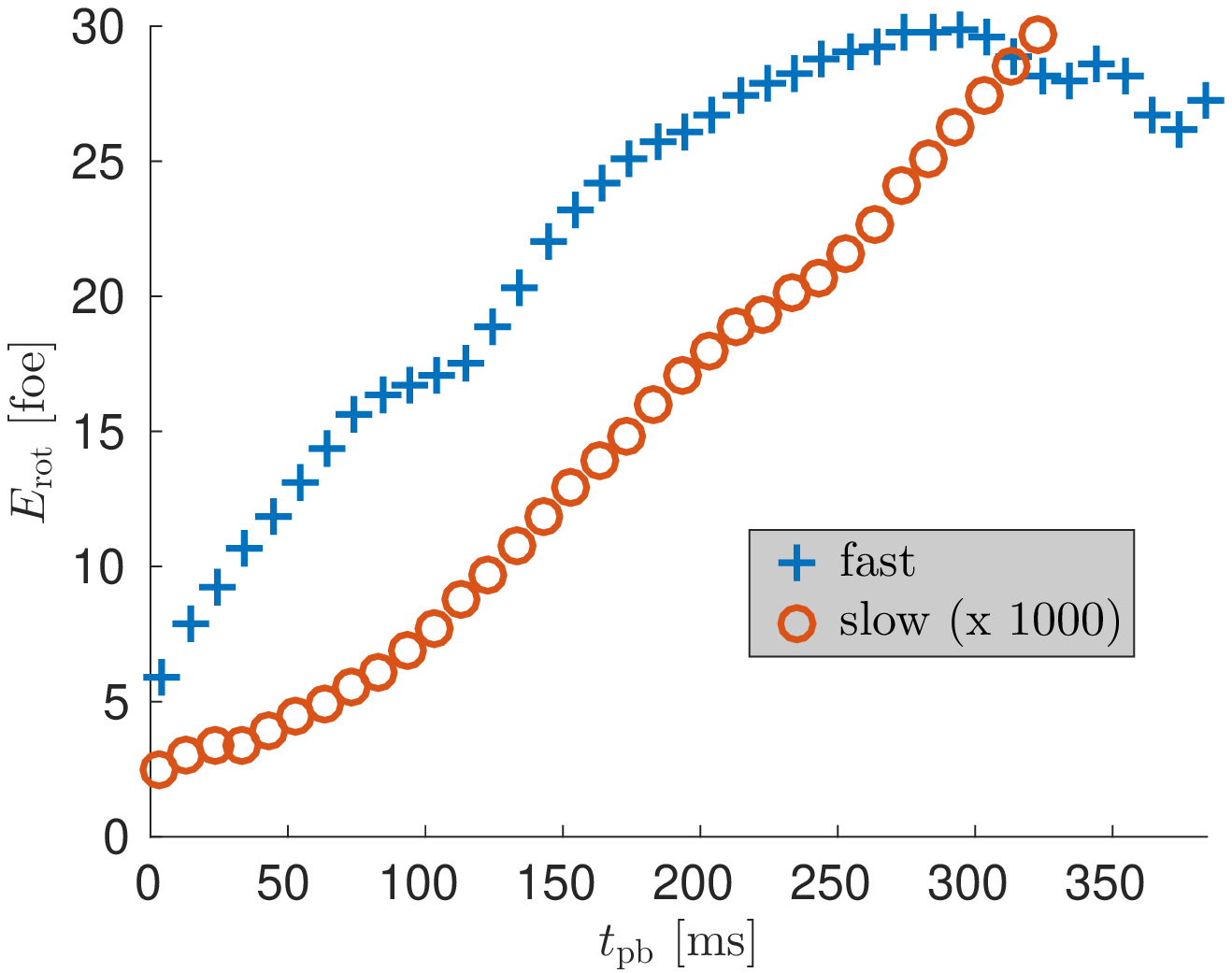} \\
    \includegraphics*[scale=0.53]{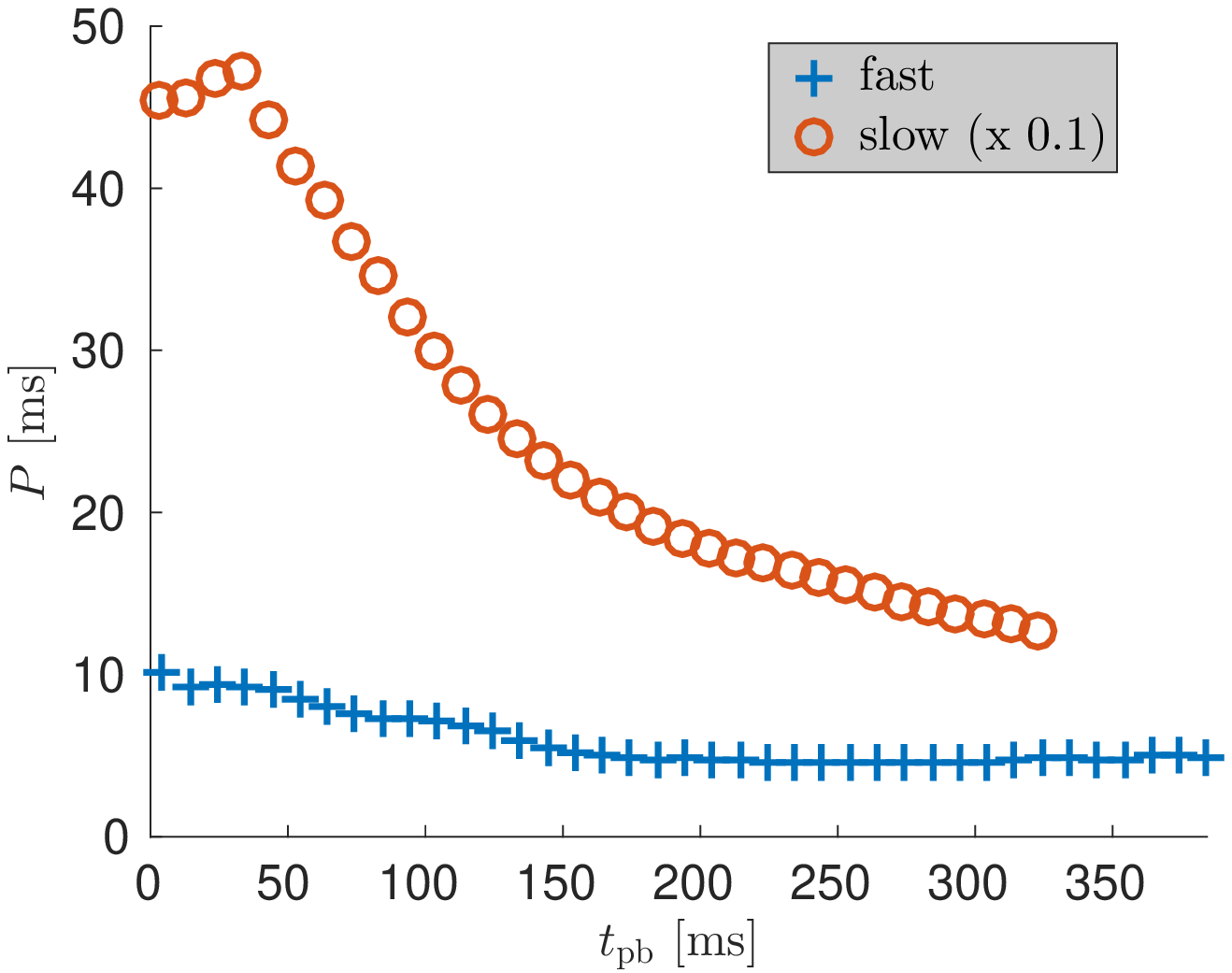} \\
    \end{tabular}
\caption{\textit{Top:} Rotational kinetic energy of the PNS
	as function of time after core bounce for the rapidly rotating progenitor (pluses)
	and for the slow rotator (circles).
	The energy of the slow rotator is multiplied by $10^3$ for the presentation.
	\textit{Bottom:} PNS spin period, where for the purpose of presentation the slow rotator period is multiplied by $0.1$ .}
      \label{fig:rot}
\end{figure}

Fig. \ref{fig:rot} shows the derived spin period and rotational kinetic energy of the PNS.
The rotation period of the PNS for the rapidly rotating case is $P_\mathrm{PNS} \approx 10 \ms$
just after core bounce,
and decreases to $P_\mathrm{PNS} \approx 5 \ms$ by the end of the simulation.
The total rotational kinetic energy grows from $E_\mathrm{rot}\approx 5 \times 10^{51} \erg$
at very early times after bounce
to $E_\mathrm{rot}\approx 3 \times 10^{52} \erg$ near the end of the simulation.
This is more than the energy required by some magnetar-driven models for SLSNe (e.g., \citealt{Chen2016}).
The PNS in the present simulation is perhaps similar to an early stage in the formation of a millisecond magnetar
which can later spin down and supply additional energy to the SN.
The details of the spin evolution are important for the energy available from magnetar spin down,
and the initial asphericity of the PNS can produce gravitational waves (e.g., \citealt{Camelio2016,Moriya2016}).
Still, for a magnetar to be relevant a successful SN explosion must first take place,
otherwise the PNS will collapse into a BH early on.
This is in accordance to the proposal of \cite{Soker2016} that jets accompany the formation of a magnetar.

\subsection{Neutron-rich disc}
\label{subsec:nucleos}

\begin{figure}
    \includegraphics*[scale=0.22]{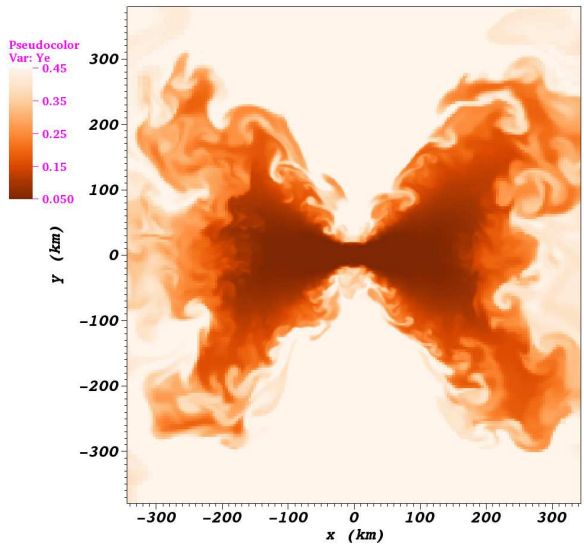} \\
      \caption{Colour map of electron fraction ($\Ye$) in the $xy$ plane
      at a post-bounce time of $\tpb=224\ms$.
      The colour coding runs from $\Ye=0.05$ (dark) 
      to $\Ye=0.45$ (light).}
      \label{fig:yedisk}
\end{figure}

Fig. \ref{fig:yedisk} shows the thick neutron-rich disc that forms around the PNS,
where the electron to nucleon ratio, $\Ye$, is relatively low.
This is due to the increased weak interactions in regions of higher density (Fig. \ref{fig:flow}).
In a thick disc-like structure with a height of several tens of kilometers,
containing a mass of approximately $0.3M_\odot$,
this ratio reaches values of $\Ye<0.1$.
\cite{Kohri2005} have previously suggested that rapid neutron capture (\textit{r}-process)
nucleosynthesis might take place due to winds from a neutron-rich disc,
in the context of a wind-driven CCSN.
If a successful explosion follows the collapse described in the simulation,
as suggested in section \ref{subsec:magnetic},
\textit{r}-process elements will be ejected into the interstellar medium.

Many uncertainties still remain regarding the sites of \textit{r}-process nucleosynthesis \citep{Thielemann2011}.
This is in particular the case for the sites of strong \textit{r}-process nucleosynthesis, where elements of atomic weight $A \ge 130$ are formed. 
According to \cite{Wehmeyer2015}, observed \textit{r}-process elements are compatible with a combined origin of
CCSNe from rapidly rotating progenitors (e.g., \citealt{Winteler2012,Nishimura2015,Nishimura2017})
and NS mergers (e.g., \citealt{Goriely2011,Hotokezaka2015,Wu2016}).
Further theoretical understanding will help in assessing the relative importance of these two types of events,
with the aid of observations such as the presence of heavy elements in stars of old dwarf galaxies \citep{Ji2016}.
\cite{Papish2015} further raised the possibility that the strong \textit{r}-process takes place in jets from a NS companion orbiting inside the core of a giant star. 
The simulation results reported in the present study strengthen the possible role of rotationally dominated CCSNe in strong \textit{r}-process nucleosynthesis.
Simulations of the long-term post-collapse evolution for numerous stellar models together with detailed nucleosynthesis calculations are needed to ascertain the amount of ejected material, its composition, and the importance for Galactic chemical evolution.
The extensive study of this issue is deferred to a future dedicated paper.

\section{DISCUSSION AND SUMMARY}
\label{sec:summary}

I have presented a study of important non-axisymmetric features in the collapse of a rapidly rotating massive star,
using the hydrodynamic code \textsc{flash}.
I have used a rapidly rotating progenitor star evolved with \textsc{mesa} excluding magnetic braking.
This allowed for the straightforward comparison to a slowly rotating case just by incorporation of the Spruit dynamo.
The outstanding shortcomings of the present study are the approximate treatment of neutrino transport,
the simulation resolution, and the somewhat artificial approach of realizing a high pre-collapse rotation rate.
Future studies will address these issues, and consider
more realistic progenitors resulting from binary interactions.

Massive stars with high core rotation rates,
as in the progenitor model used in this study,
might be relatively rare as they need strong binary interaction to acquire their angular momentum.
The collapse and possible explosion of such stars can still be significant in explaining some observed phenomena.
A prolonged bipolar outflow, as suggested in section \ref{subsec:magnetic},
can function as the engine of extremely energetic SNe.
This agrees with the elongated morphology implied by recent observations \citep{Inserra2016}.
Asymmetric momentum distribution (section \ref{subsec:natal}) can give birth to BHs with significant natal kicks.
The neutron-rich disc discussed in section \ref{subsec:nucleos} can induce strong \textit{r}-process nucleosynthesis of heavy elements.

A rapidly rotating strongly magnetized PNS formed in a similar way to the presented simulation can be an early-stage millisecond magnetar,
which might later deposit a large amount of energy in the SN ejecta during its spin-down.
First, an explosion must take place.
For progenitors which are very tightly bound gravitationally,
an explosion is unlikely to be driven by the neutrino flux alone, whereas a jet-driven SN might be more promising.
Estimating the quantitative energy contributions of the jets and the magnetar requires further study,
yet it seems the energy available should suffice for SLSNe.
I suggest that Type I SLSNe result from flow dynamics qualitatively similar to the presented simulation,
and also contribute to the production of strong \textit{r}-process elements.

\section*{Acknowledgments}

I thank Noam Soker for his guidance and for fruitful discussions,
and an anonymous referee for providing constructive comments which helped improve the manuscript.
The software used in this work was developed in part by the DOE NNSA-ASC
and DOE Office of Science ASCR-supported Flash Center for Computational Science at the University of Chicago.
Simulations were run on the Israeli astrophysics I-CORE astric HPC.
The \textsc{yt} package \citep{Turk2011} was extensively used for analysis of the results.
The author is supported by the Blavatnik Family Foundation.

\appendix
\section{Effect of magnetic braking on the pre-collapse core rotation rate}
\label{app:corej}

The stellar models used for the present study are taken from a moderately broad set of models evolved in the same approach,
with different initial conditions.
The models vary in their initial mass, from $12 M_\odot$ to $90 M_\odot$, with logarithmic spacing.
The initial rotation rates are taken between $10\%$ and $90\%$ of the breakup velocity.

\begin{figure}
    \includegraphics*[scale=0.34]{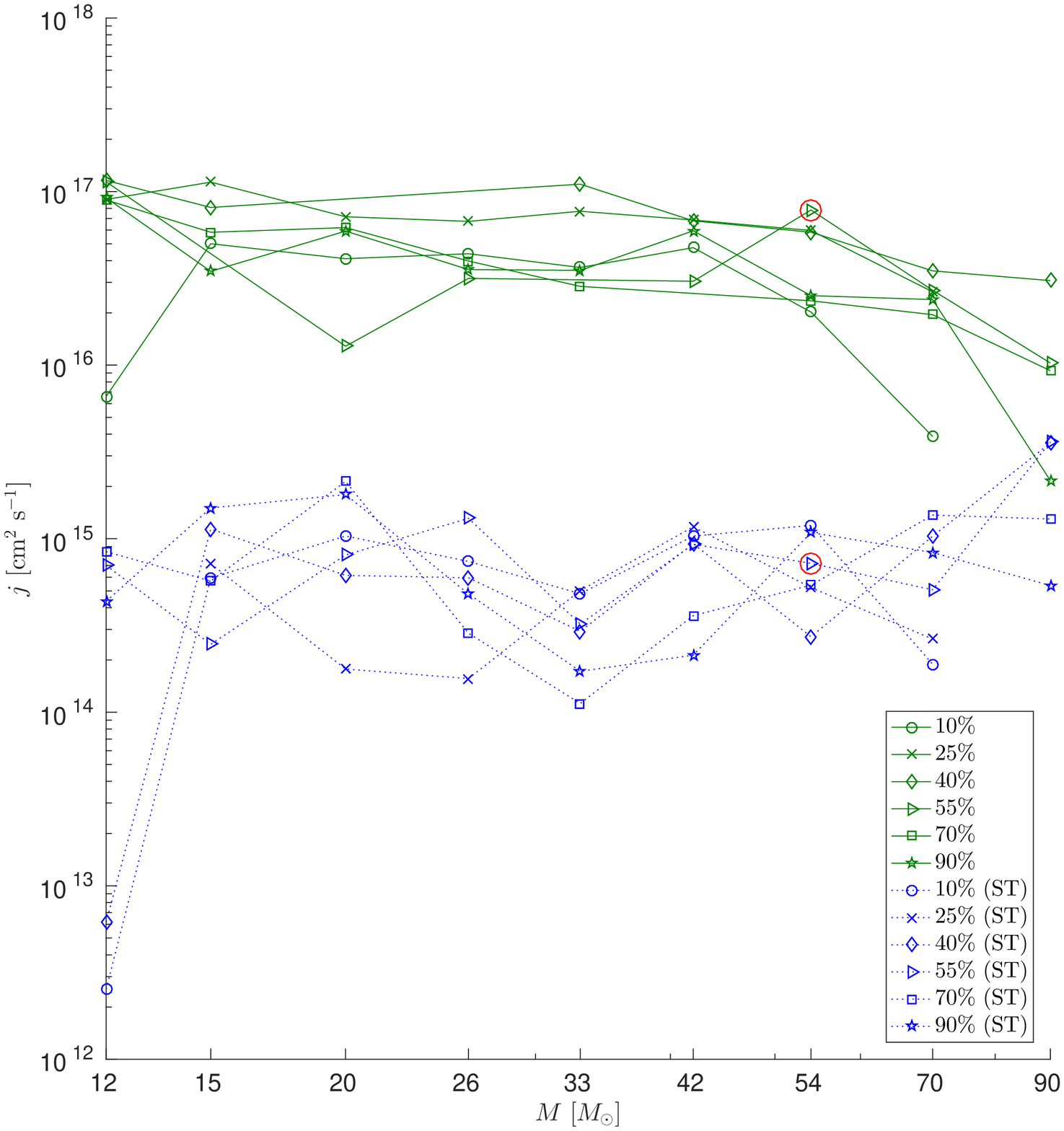} \\
\caption{The specific angular momentum at the outer edge of the iron core,
for 98 stellar models. The models differ by their initial mass,
initial rotation rate, and inclusion of the Spruit dynamo. Red circles denote the models used in the present study.}
      \label{fig:j}
\end{figure}

Fig. \ref{fig:j} succinctly summarizes this set of models,
showing just the specific angular momentum at the outer edge of the iron core.
The value of the specific angular momentum at the core edge is mostly concentrated in the range
$10^{14} \cm^2 \s^{-1} \la j_\mathrm{iron} \la 10^{15} \cm^2 \s^{-1}$ for models where the Spruit dynamo is taken into account,
and $10^{16} \cm^2 \s^{-1} \la j_\mathrm{iron} \la 10^{17} \cm^2 \s^{-1}$ when the dynamo is neglected.
It is clearly seen that disregarding magnetic braking results in far higher core rotation rates.
This is in agreement with the results of \cite{Heger2005},
for a wider range of initial parameters.
Moreover, the value of the specific angular momentum at the core edge is for the most part unaffected by the initial mass and the initial rotation rate.

\section{Simulations at higher resolutions}
\label{app:res}

\begin{figure}
\begin{tabular}{c}
    \includegraphics*[scale=0.22]{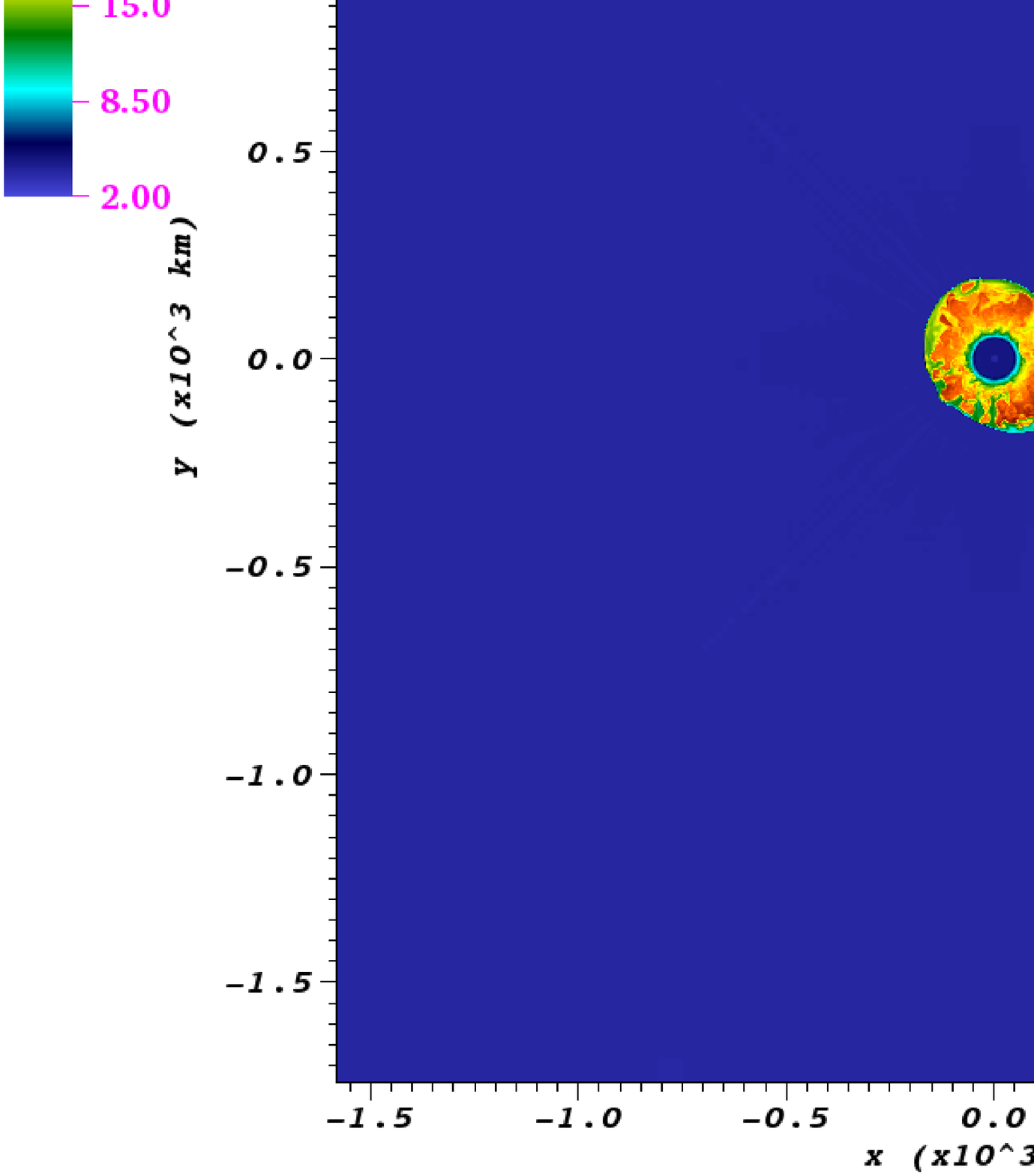} \\
    \includegraphics*[scale=0.22]{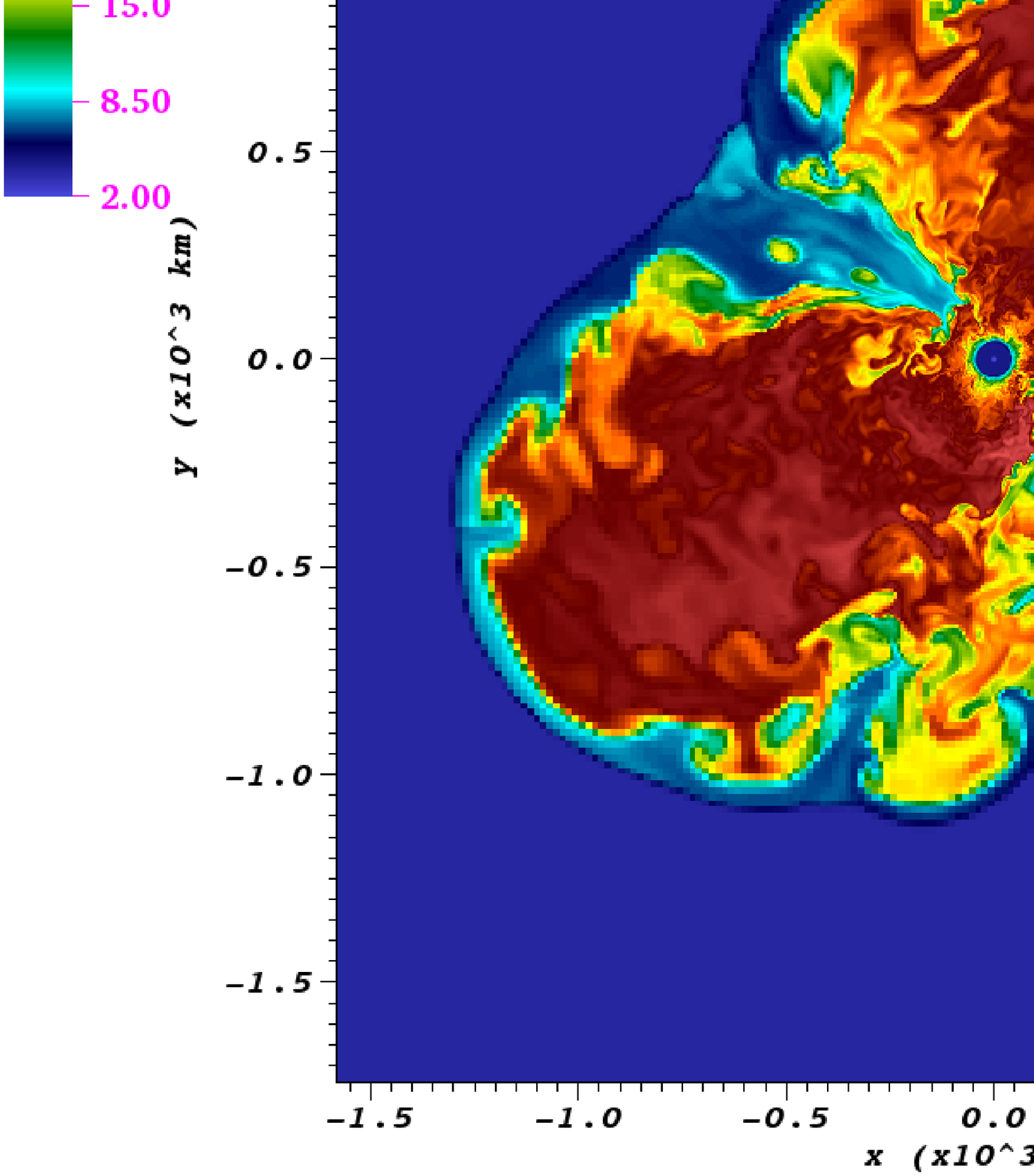} \\
\end{tabular}
	\caption{Entropy colour maps in the $xy$ plane,
      where $y$ is the axis of rotation, at $\tpb=323\ms$ for the simulation of the slowly rotating progenitor with finest resolutions of $\Delta x\simeq 1.95 \km$ (top) and $\Delta x\simeq 0.98 \km$ (bottom). Entropy is shown in units of $k_\mathrm{B}$ per nucleon.}
      \label{fig:slowentropy}
\end{figure}

\begin{figure}
    \includegraphics*[scale=0.53]{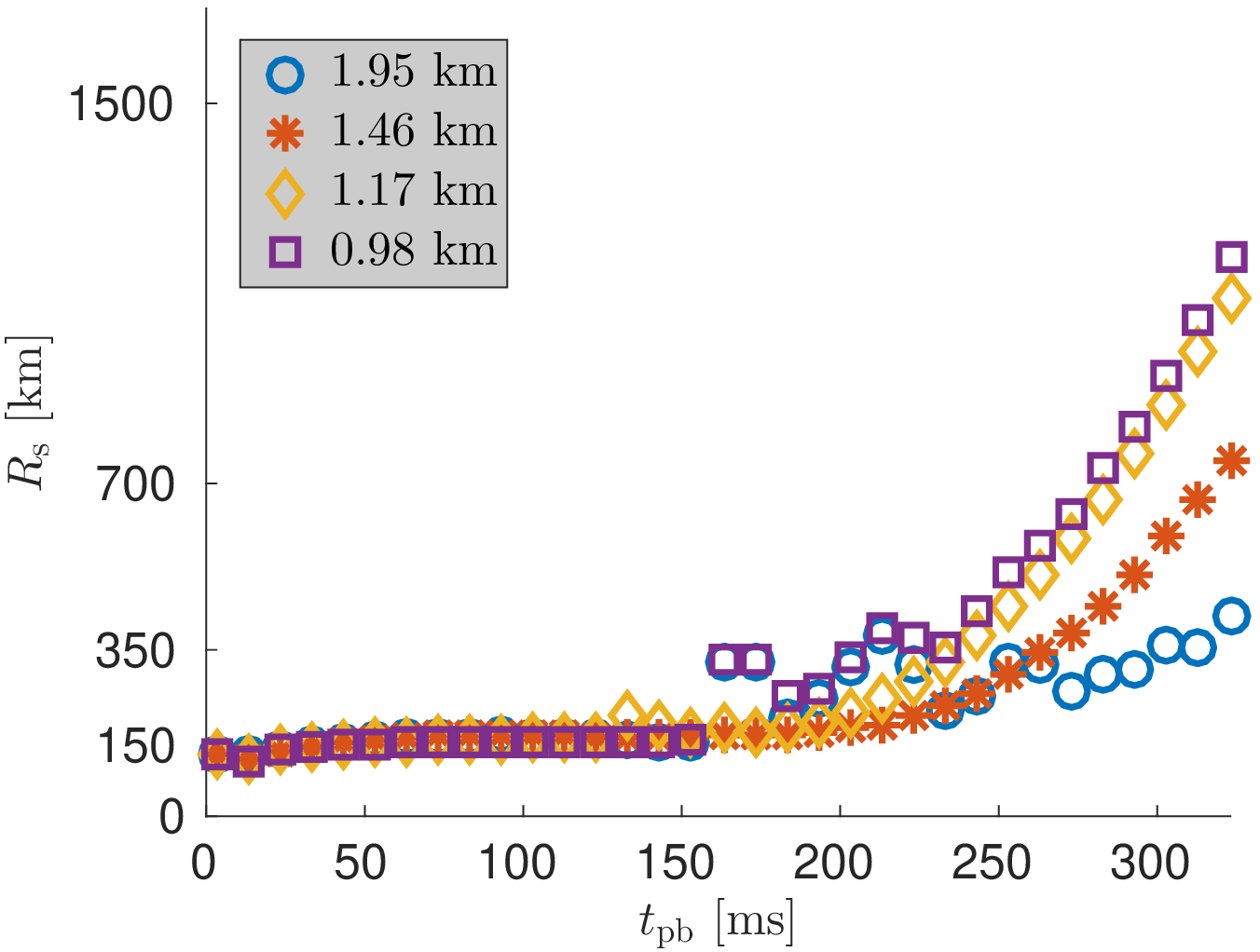} \\
\caption{Shock radius calculated from the volume contained within the surface on which the entropy per nucleon is $6k_\mathrm{B}$,
as function of time after core bounce for simulations with different finest resolutions.}
      \label{fig:slowshock}
\end{figure}

Additional simulations were run to check the sensitivity to the grid resolution,
with finest resolutions refined by factors of $0.75$, $0.6$ and $0.5$.
The simulations of the slowly rotating progenitor were all run for $600\ms$.
Fig. \ref{fig:slowentropy} show the flow structure at the end of the simulation,
for two simulations of the slowly rotating progenitor, with finest resolutions of $1.95\km$ and $0.98\km$.
The expansion of the high-entropy region is greatly increased for the refined simulation,
as shown also in Fig. \ref{fig:slowshock}.

\begin{figure}
    \includegraphics*[scale=0.53]{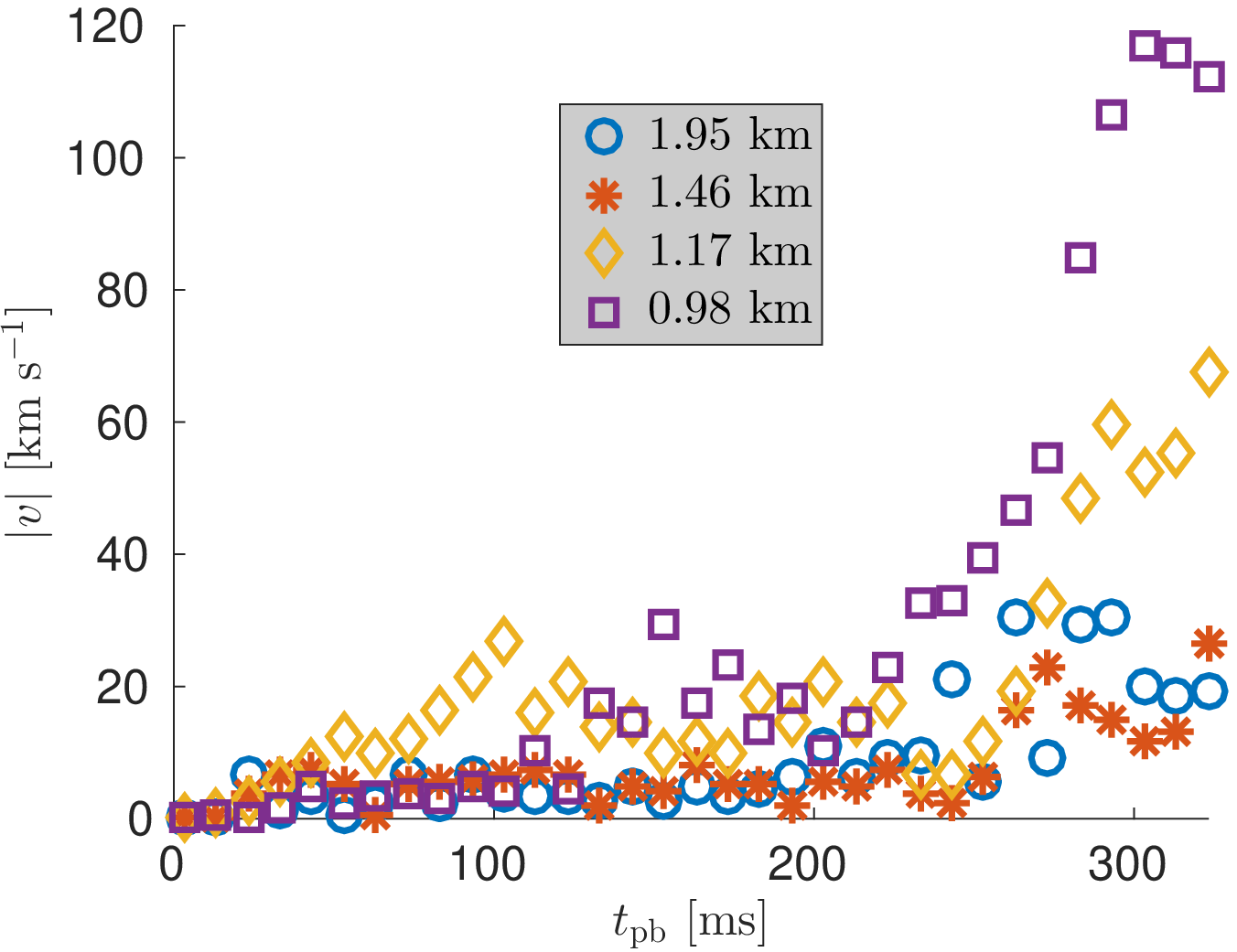} \\
      \caption{Magnitude of the average velocity of the PNS for the slowly rotating progenitor, for simulations with different finest resolutions.}
      \label{fig:slowkick}
\end{figure}

\begin{figure*}
\begin{tabular}{cc}
    \includegraphics*[scale=0.53]{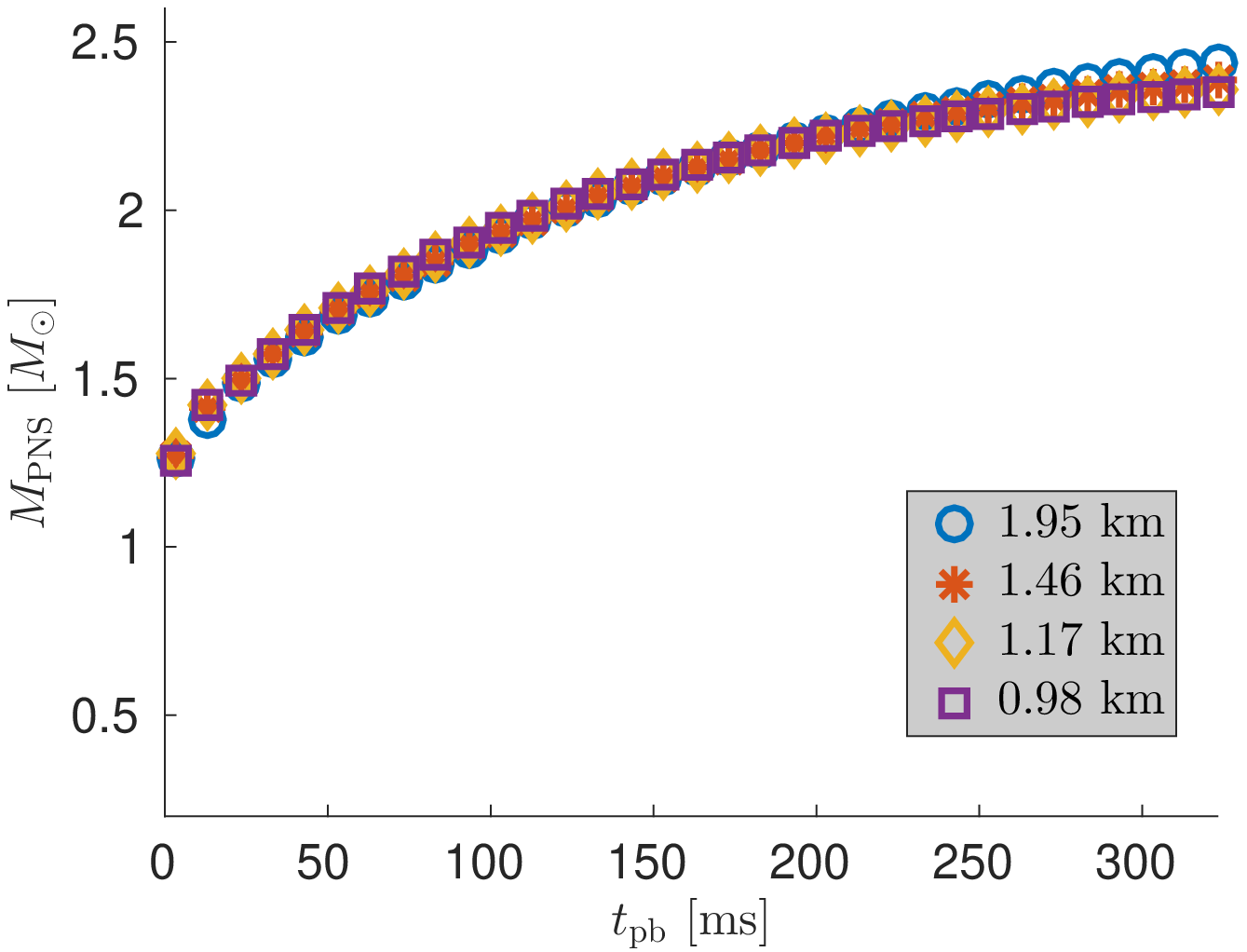} &
    \includegraphics*[scale=0.53]{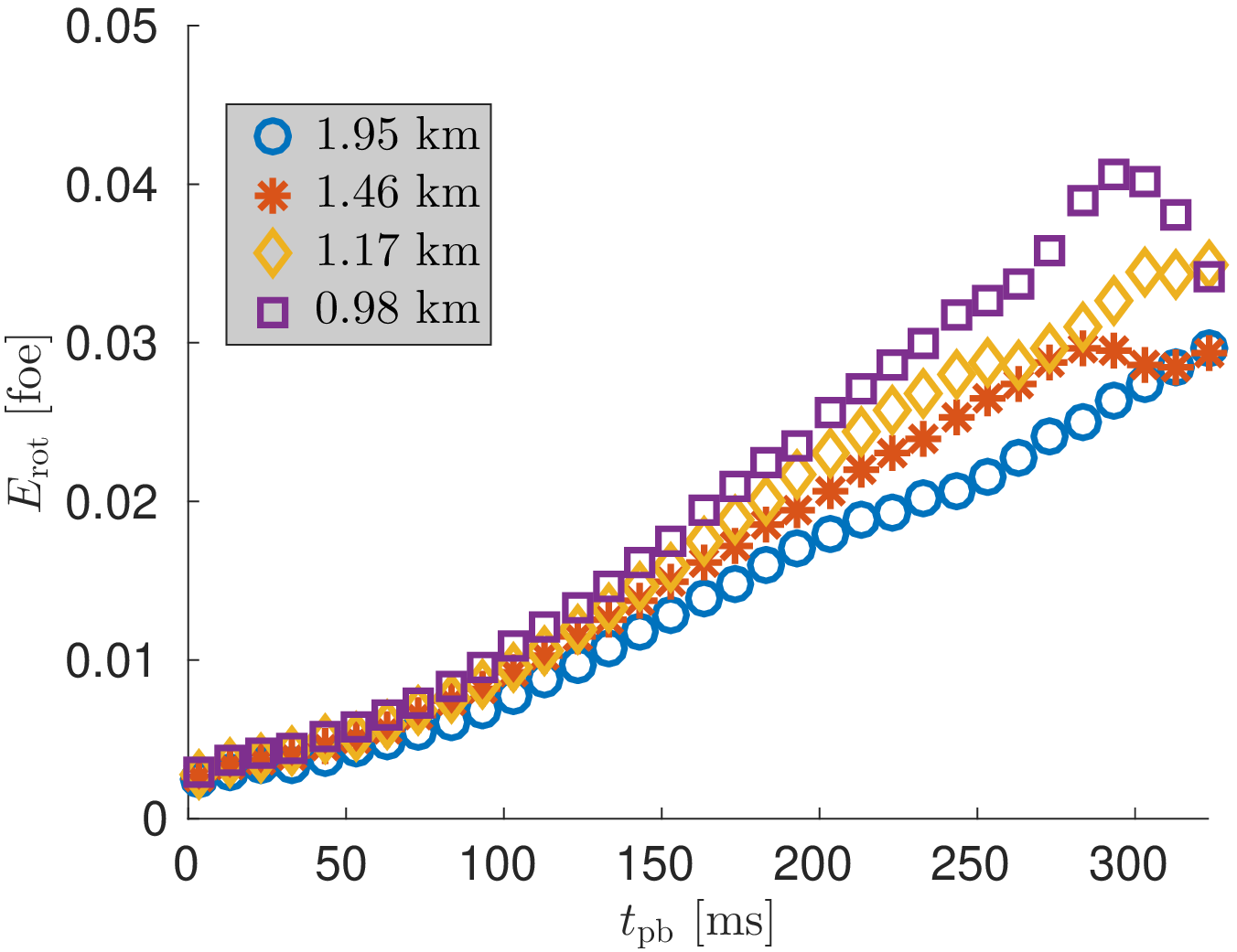} \\
    \includegraphics*[scale=0.53]{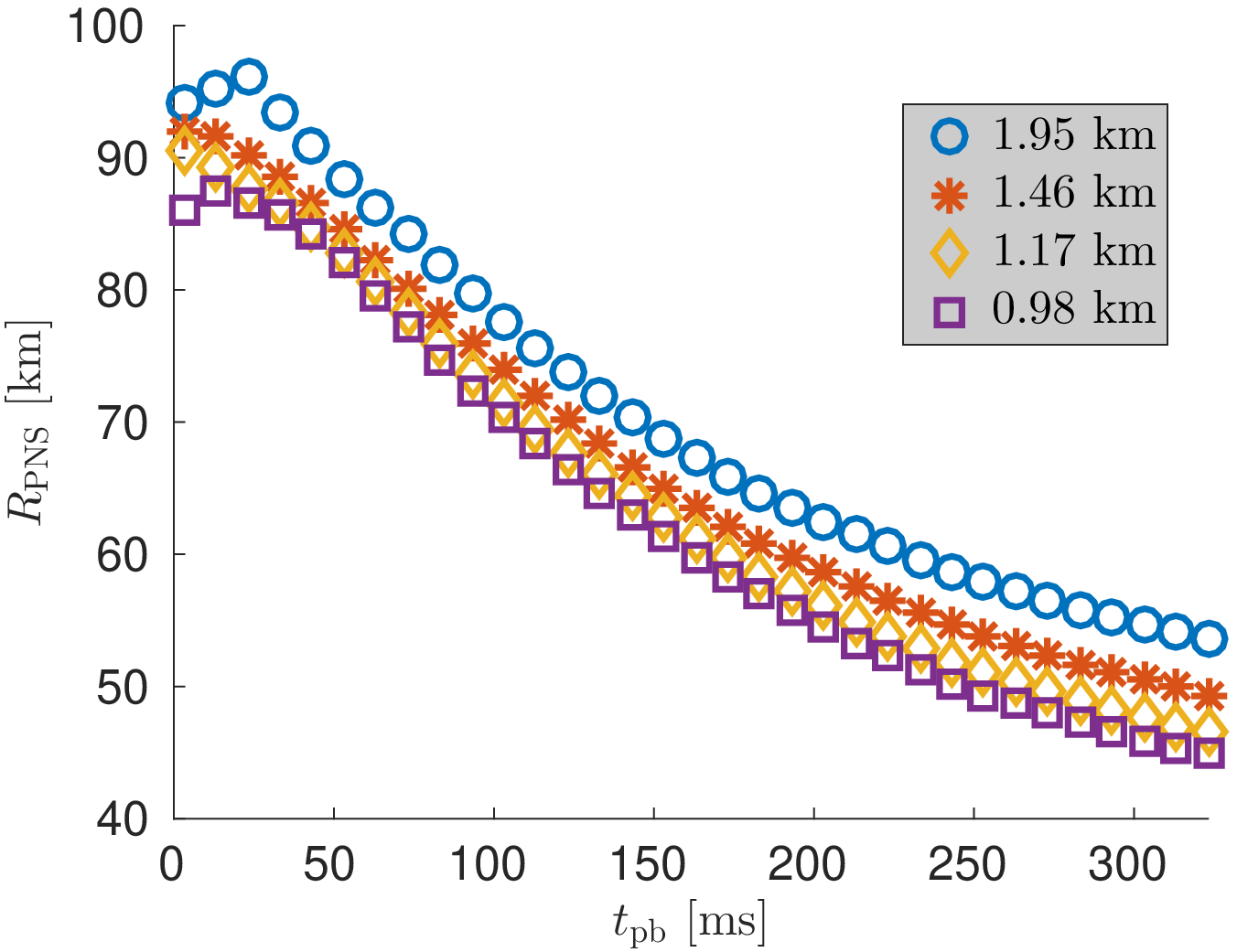} &
    \includegraphics*[scale=0.53]{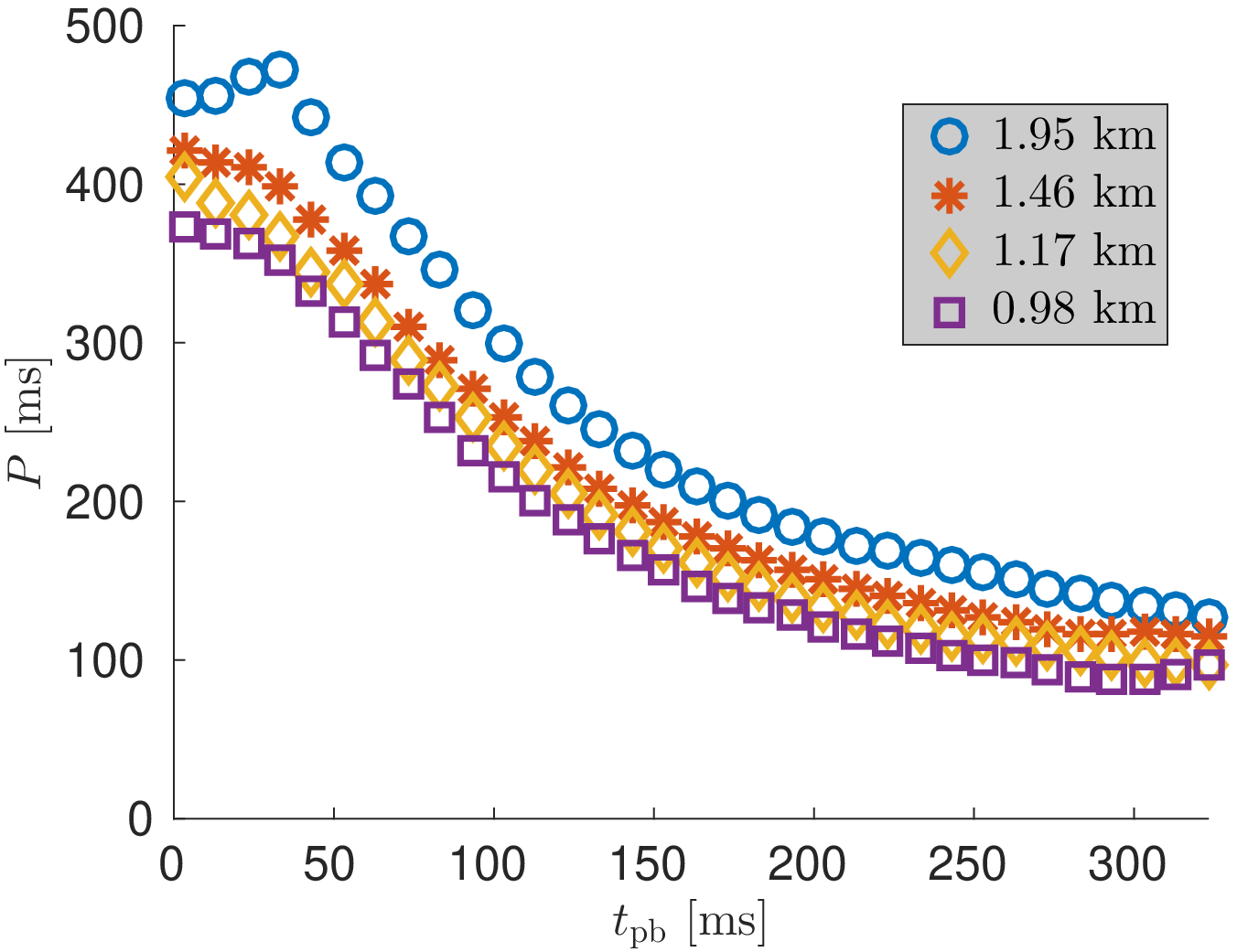} \\
    \end{tabular}
\caption{\textit{Top-left:} Baryonic PNS mass as function of time after core bounce for the slowly rotating progenitor,
	for simulations with different finest resolutions.
	\textit{Top-right:} Rotational kinetic energy of the PNS.
	\textit{Bottom-left:} PNS effective radius.
		\textit{Bottom-right:} PNS spin period.}
      \label{fig:slowpns}
\end{figure*}

Higher resolution seems to increase the diagnostic energy as well,
with the refined simulation reaching $E_\mathrm{diag}\simeq 1.1 \foe$.
The derived kick is accordingly stronger (Fig. \ref{fig:slowkick}),
but still lower than the highest values obtained in the nominal rapidly rotating simulation.
The other PNS parameters are mostly insensitive to the simulation resolution (Fig. \ref{fig:slowpns}),
although there are a few differences.
For example, the PNS becomes distinctly more compact with increased refinement,
with the rotation period decreasing accordingly (similar to its correlation with decreasing radius).
The particular numerical details which cause this change with resolution are unclear,
although it is evident that the effects of resolution must be checked for such calculations.

\begin{figure}
\begin{tabular}{c}
    \includegraphics*[scale=0.53]{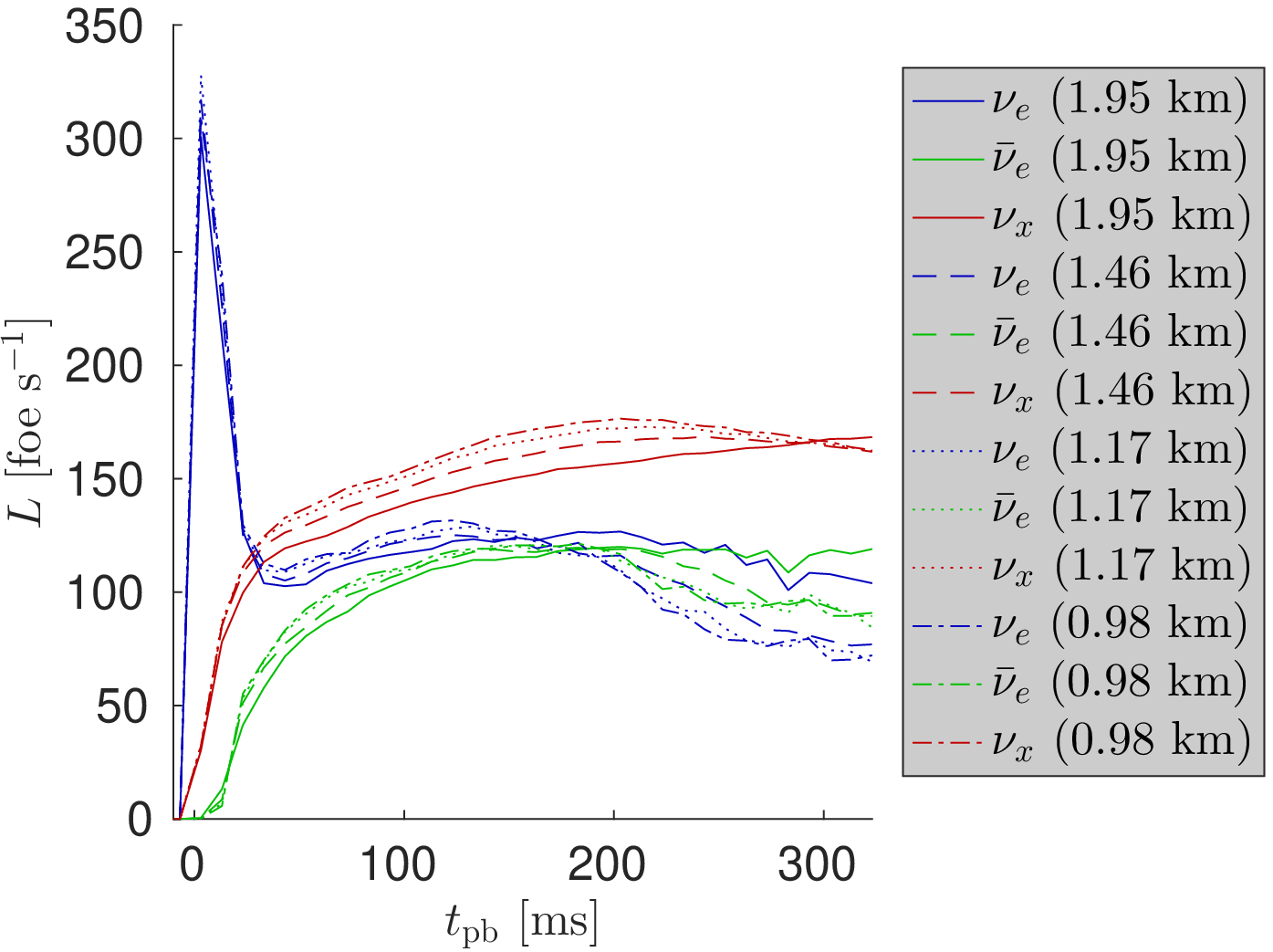} \\
    \includegraphics*[scale=0.53]{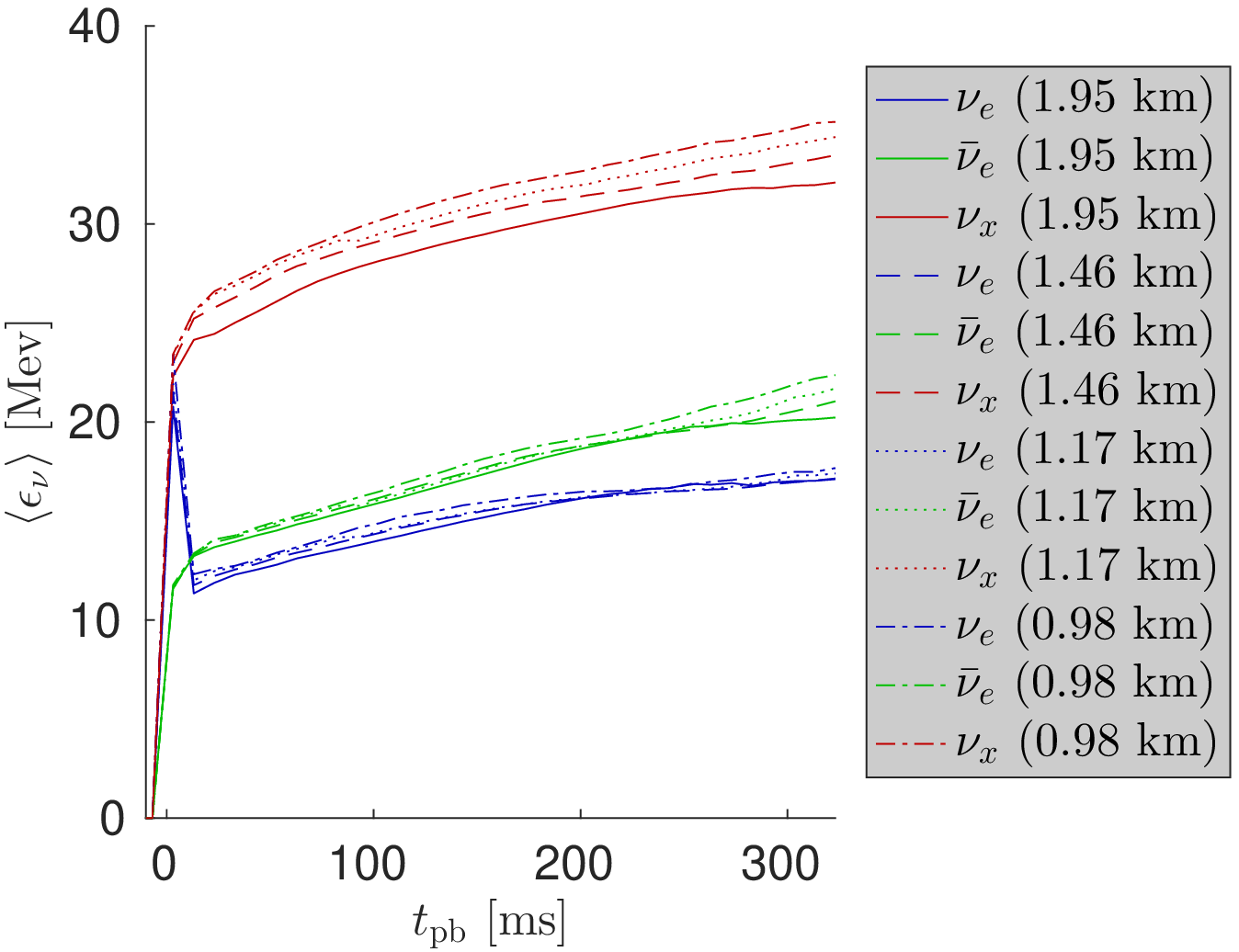} \\
    \end{tabular}
\caption{\textit{Top:} Luminosity of the neutrino luminosity for each neutrino species as function of time after core bounce for the slowly rotating progenitor, for simulations with different finest resolutions.
		\textit{Bottom:} Average neutrino energy in the simulations.}
      \label{fig:neu}
\end{figure}

The details of the neutrino heating are presented in Fig. \ref{fig:neu}. The average neutrino energy increases for the more refined simulations, corresponding to the more contracted PNS.
The neutrino luminosity also shows a clear sequence, increasing with resolution refinement, up until $\tpb \la 200 \ms$.
The sequence is then broken, as the contribution of accreting matter vanishes.

\begin{figure}
    \includegraphics*[scale=0.53]{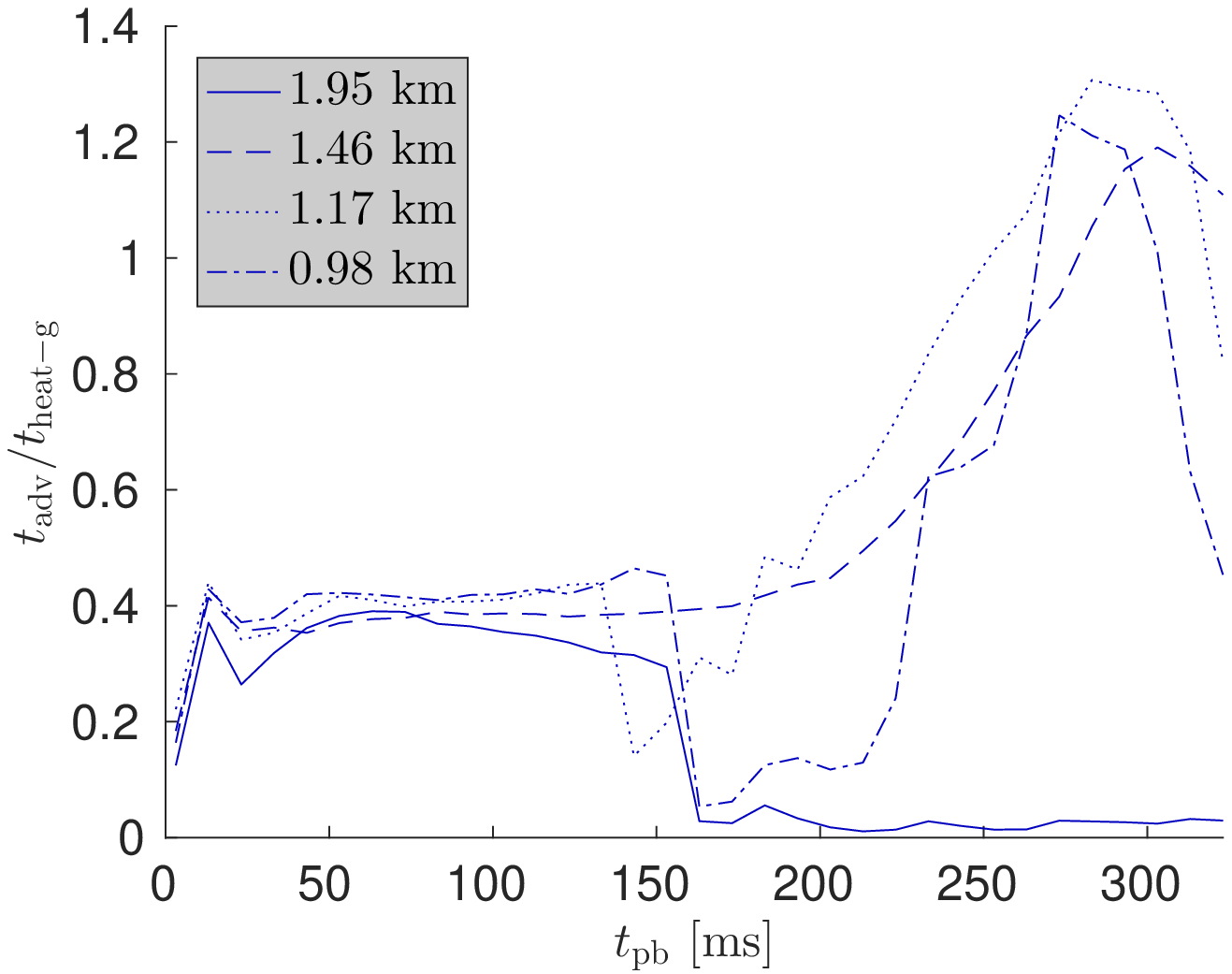} \\
\caption{Ratio of the advection time to the heating time in the gain region as function of time after core bounce for the slowly rotating progenitor, for simulations with different finest resolutions.}
      \label{fig:timescales}
\end{figure}

Fig. \ref{fig:timescales} presents the ratio of advection time to heating time in the gain region.
The advection time is calculated by $t_\mathrm{adv}=\frac{\langle R_\mathrm{s} \rangle - \langle R_\mathrm{g} \rangle}{\langle v_r \rangle}$ (e.g., \citealt{MarekJanka2009}),
where $\langle R_\mathrm{s} \rangle$ is the average radius of the shock, $\langle R_\mathrm{g} \rangle$ is the average inner radius of the gain region (where material is heated by neutrinos), and $\langle v_r \rangle$ is the mass-weighted average radial velocity in the gain region.
The heating time is simply the ratio of the total internal energy to the net neutrino heating rate (e.g., \citealt{Fernandez2012}).
This ratio shows the effect of hydrodynamic motion on the neutrino heating,
and here the greatest discrepancy is seen between the coarsest simulation and the more refined simulations.

Simulations of the rapidly rotating progenitor proved much more difficult than those of the slowly rotating progenitor.
For the nominal simulation, the time step had to be limited by a factor of $0.1$ relative to the Courant-Friedrichs-Lewy (CFL) condition to avoid numerical non-convergence,
whereas for the slow rotator a CFL factor of $0.5$ was applied throughout.
Moreover, the refined simulations of the rapidly rotating progenitor all terminated around $\tpb \approx 60 \ms$ due to numerical problems,
even with a CFL factor of $0.1$.
Up to this time,
the PNS parameters evolve similarly for all resolutions.
It is expected that as for the slow rotator,
here as well the PNS will be insensitive to resolution,
while the high-entropy region surrounding it will be affected.

\ifmnras
	\bibliographystyle{mnras}

\label{lastpage}

\end{document}